\documentclass[journal=jctcce,manuscript=article,email=false]{achemso}
\setkeys{acs}{articletitle=true,etalmode=truncate,maxauthors=0}

\usepackage[colorlinks=true,linkcolor=blue,citecolor=blue,urlcolor=blue]{hyperref}
\usepackage{setspace}
\usepackage{graphicx}
\usepackage{amsmath}
\usepackage{color}
\usepackage{amssymb}
\usepackage{verbatim}
\usepackage{latexsym}
\usepackage{enumerate}
\usepackage{bm}
\usepackage{caption}
\usepackage{subcaption}
\usepackage{mathtools}
\usepackage{braket}
\usepackage{multirow}
\usepackage[T1]{fontenc}
\usepackage{mciteplus}
\usepackage{xr}

\title{Efficiently computing excitations of complex systems: linear-scaling time-dependent embedded mean-field theory in implicit solvent}

\author{Joseph C.\ A.\ Prentice}
\affiliation{Department of Materials, University of Oxford, Parks
  Road, Oxford OX1 3PH, United Kingdom}

\begin{document}

\begin{abstract}

  Quantum embedding schemes have the potential to significantly reduce
  the computational cost of first principles calculations, whilst
  maintaining accuracy, particularly for calculations of electronic
  excitations in complex systems. In this work, I combine
  time-dependent embedded mean field theory (TD-EMFT) with
  linear-scaling density functional theory and implicit solvation
  models, extending previous work within the \textsc{onetep}
  code. This provides a way to perform multi-level calculations of
  electronic excitations on very large systems, where long-range
  environmental effects, both quantum and classical in nature, are
  important. I demonstrate the power of this method by performing
  simulations on a variety of systems, including a molecular dimer, a
  chromophore in solution, and a doped molecular crystal. This work
  paves the way for high accuracy calculations to be performed on
  large-scale systems that were previously beyond the reach of quantum
  embedding schemes.
  
\end{abstract}

\section{Introduction} \label{sec:Introduction}

Embedding schemes are a well-studied method for improving the
computational efficiency of calculations on complex systems, without
significantly sacrificing accuracy. These schemes are best suited to
systems where the relevant physics is dominated by a small `active'
sub-region, but the rest of the system still affects this behaviour on
an environmental level\cite{sun_quantum_2016}. In such systems, a
certain level of theory may be required to accurately describe the
relevant physics, but applying this level of theory to the whole
system is often infeasibly computationally expensive. Examples could
include molecules in
solution\cite{altun_spectral_2008,isborn_electronic_2012,zuehlsdorff_solvent_2016},
host-guest
systems\cite{eisbein_proton_2014,hirao_multiscale_2015,witman_rational_2017},
defects in
crystals\cite{huber_qm/mm_2016,chen_qm/mm_2016,chen_qm/mm_2017}, and
active sites in
enzymes\cite{mulholland_chemical_2007,cole_applications_2016,kulik_large-scale_2018}. Embedding
schemes seek to solve this problem by treating the active region with
an accurate, but computationally intensive, `higher' level of theory,
whilst the environment is treated with a less demanding, but less
accurate, `lower' level of theory. Using the higher level of theory
for the active region only means that the most important contributions
to the property under study are still described accurately, whilst
using the lower level of theory for the rest of the system reduces the
computational cost, but still allows the environment to influence the
result.

Embedding schemes can be divided into those that treat the environment
classically\cite{warshel_theoretical_1976,svensson_oniom:_1996}, and
those that treat the environment quantum
mechanically\cite{wesolowski_frozen_1993,staroverov_optimized_2006,goodpaster_exact_2010,culpitt_communication:_2017,huzinaga_theory_1971,manby_simple_2012,hegely_exact_2016},
allowing for quantum mechanical interactions between the
regions\cite{sun_quantum_2016}; the latter class are known as quantum
embedding schemes. One recently proposed such scheme is embedded
mean-field theory (EMFT)\cite{fornace_embedded_2015}. Among other
advantages over other quantum embedding schemes, EMFT is a mean-field
theory, like density functional theory (DFT), so many existing methods
that have been built on the foundations of DFT can be easily modified
to accommodate EMFT. EMFT has been successfully used several times
since its
proposal\cite{miyamoto_fock-matrix_2016,ding_embedded_2017,ding_linear-response_2017,koh_accelerating_2017,jiang_imaging_2019,chen_embedded_2020},
largely focused on relatively small molecular systems. In a previous
publication\cite{prentice_combining_2020}, however, the author and
co-workers extended the applicability of EMFT to large-scale periodic
systems by presenting a novel combination of EMFT and linear-scaling
DFT in the code \textsc{onetep}\cite{prentice_onetep_2020}. This work
demonstrated EMFT's utility for hybrid DFT-in-semi-local DFT embedding
calculations on large-scale systems, such as molecular crystals, but
focused on calculating ground state energies only. Although we were
able to access some excited state properties, studying the excited
states of such large systems more generally with EMFT was not
considered.

One of the most popular methods for calculating electronic excitations
is time-dependent density functional theory (TDDFT). TDDFT is popular
for its balance of reasonable accuracy and relatively low
computational
cost\cite{kronik_excited-state_2016,zuehlsdorff_linear-scaling_2013}. However,
standard semi-local TDDFT has several known issues, including its
failure to correctly describe charge transfer
states\cite{zuehlsdorff_linear-scaling_2013,peach_excitation_2008},
and the underestimation of excitation
frequencies\cite{zuehlsdorff_solvent_2016}. These issues can be
partially fixed by using hybrid functionals, including range-separated
hybrids, but these are significantly more computationally
expensive\cite{peach_excitation_2008}. Quantum embedding offers a way
to obtain the accuracy of these methods, whilst significantly lowering
the computational cost. The combination of linear-response TDDFT and
EMFT (known as TD-EMFT) has previously been implemented and found to
work well, but has only been applied to small molecular
systems\cite{ding_linear-response_2017}.

In this work, I extend the previously described novel combination of EMFT and
linear-scaling DFT to include TD-EMFT, allowing electronic excitations
to be computed using this scheme. I also combine this
implementation with the implicit solvation model present in
\textsc{onetep}\cite{dziedzic_minimal_2011,prentice_onetep_2020},
allowing for both EMFT and TD-EMFT calculations to be placed in a
continuous dielectric medium with a given permittivity. This makes
multi-level calculations of electronic excitations possible -- for
example, using a hybrid functional to describe the active region, a
semi-local function to describe the nearby environment at a quantum
level, and then implicit solvent to describe the rest of the
environment at a continuum level. This allows for computationally
efficient and highly accurate TDDFT calculations to be performed on
much larger systems than would previously have been possible. I have
tested this implementation on a range of different systems,
demonstrating the breadth of potential applications.

The work is organised as follows. In Section \ref{sec:Theory}, I give
a brief overview of the theory of (TD-)EMFT as described in previous
work, and how this is implemented in \textsc{onetep}. In Section
\ref{sec:Examples}, I give the results of testing our linear-scaling
TD-EMFT implementation on several systems: a water-nitrogen dimer
(Sec.\ \ref{subsec:Dimer}), phenolphthalein solvated in water
(Sec.\ \ref{subsec:PhPh}), and a pentacene-doped \textit{p}-terphenyl
molecular crystal (Sec.\ \ref{subsec:PentPTer}). Finally, in Section
\ref{sec:Conclusions}, I give some concluding remarks.

\section{Background theory} \label{sec:Theory}

In this work, atom-centered basis functions are used, which in general
will be non-orthogonal. Because of this, the overlap matrix $S$, which
gives the overlaps between basis functions, acts as a metric tensor in
the space spanned by the basis functions. As $S$ is not simply the
identity in general, a distinction must be drawn between covariant and
contravariant quantities, represented with subscript and superscript
indices in the following. A contravariant quantity $\chi^\beta$ can be
transformed into its dual covariant quantity $\chi_\alpha$ by applying
$S$: $\chi_\alpha = \sum_\beta S_{\alpha \beta} \chi^\beta$, where
$\alpha, \beta$ run over basis functions. Conversely, covariant
quantities can be transformed into their dual contravariant quantities
using the inverse overlap $S^{-1}$: $\chi^\beta = \sum_\alpha
\left(S^{-1}\right)^{\beta \alpha} \chi_\alpha$. Greek indices are
used to enumerate the basis functions, with capital Latin indices
representing different embedding regions.

\subsection{Ground-state embedded mean-field theory} \label{subsec:EMFT}

As outlined in previous work, EMFT is based on splitting the system
into two regions -- the active region $A$ and the environment $B$ --
at the basis set level\cite{fornace_embedded_2015}. For atom-centered
basis sets, this simply means assigning each atom to a particular
region, which then assigns all basis functions associated with that
atom to that region too. If the density matrix is expressed in terms of
these basis functions (also known as the density kernel
$K$\cite{prentice_onetep_2020}), it can be separated into blocks
corresponding to the regions:
\begin{equation}
  K = \begin{pmatrix} K^{AA} & K^{AB} \\
    K^{BA} & K^{BB} \end{pmatrix} ~. \label{eq:EmbedK}
\end{equation}
A similar expression applies for the overlap matrix $S$. The density
of the full system can then be calculated as $\rho(\mathbf{r}) =
\sum_{\alpha\beta} \phi_{\alpha}^*(\mathbf{r}) K^{\alpha\beta}
\phi_{\beta}(\mathbf{r})$, where $\phi_{\alpha}(\mathbf{r})$ are the
basis functions. Densities corresponding to the various blocks of $K$
can be calculated as $\rho_{IJ}(\mathbf{r}) = \sum_{\alpha\in I,
  \beta\in J} \phi_{\alpha}^*(\mathbf{r}) K^{\alpha\beta}
\phi_{\beta}(\mathbf{r})$.

The energy can now be written as a functional of $K$ in its most
general form for a mean-field theory, as EMFT is only applicable to
mean-field theories. The energy is given
by\cite{fornace_embedded_2015,prentice_combining_2020}
\begin{equation}
  E[K] = E_{\text{1-el}}[K] + E_{\text{2-el}}[K] ~,
  \label{eq:EMFTGeneralE}
\end{equation}
where $E_{\text{1-el}}$ corresponds to the energy arising from all
one-electron terms in the Hamiltonian, and $E_{\text{2-el}}$
corresponds to the energy arising from all two-electron terms. In DFT,
$E_{\text{1-el}}$ includes contributions such as the kinetic and
electron-nuclear contributions to the energy, whilst $E_{\text{2-el}}$
includes the Hartree and exchange-correlation contributions.

As this work focuses on DFT-in-DFT embedding, the higher and lower
levels of theory can be assumed to differ only in the two-electron
term -- the higher level of theory would have $E^{\text{high}}[K] =
E_{\text{1-el}}[K] + E_{\text{2-el}}^{\text{high}}[K]$, whilst the
lower level would have $E^{\text{low}}[K] = E_{\text{1-el}}[K] +
E_{\text{2-el}}^{\text{low}}[K]$. The key assumption of EMFT is then
that the energy can be written as
as\cite{fornace_embedded_2015,prentice_combining_2020}
\begin{equation}
  E^{\text{EMFT}}[K] = E_{\text{1-el}}[K] +
  E_{\text{2-el}}^{\text{low}}[K] + \left( E_{\text{2-el}}^{\text{high}}[K^{AA}] -
  E_{\text{2-el}}^{\text{low}}[K^{AA}] \right) ~. \label{eq:EMFTEnergy}
\end{equation}
Three energy evaluations are required to evaluate this expression:
firstly, the energy of the whole system (including the one-electron
terms) is calculated at the lower level of theory. Next, the
two-electron terms are computed twice using the $K^{AA}$ sub-block of
the density kernel only -- once at the lower level of theory, and once
with the higher level. The difference of these two quantities is
calculated, and added on as a correction to the energy of the whole
system calculated previously. All the quantities computed here are
calculated at the mean-field level, so this is a mean-field theory.

In DFT-in-DFT embedding, $E_{\text{2-el}}$ depends on the
exchange-correlation functional chosen. The most logical choice for
the lowest level of theory is to use a semi-local functional, with the
higher level of theory using a more computationally demanding type of
functional, such as a hybrid functional. Importantly, hybrid
functionals include a fraction of exact exchange energy. Exact
exchange, unlike the other energy terms discussed so far, is not a
functional of the density, so needs to be treated differently. The
least computationally expensive way of calculating the exact exchange
contribution to the energy of the active region is the EX0
method\cite{fornace_embedded_2015}. This only includes exchange within
the active region, neglecting exchange between the active region and
the environment. It is possible to include exchange between the active
region and the environment, but previous work has shown that this does
not significantly improve accuracy, and also increases computational
cost\cite{ding_embedded_2017}. The EX0 method for exact exchange is
therefore used throughout this work.

Previous work has also shown that in many situations, a block
orthogonalisation procedure is required to prevent an EMFT calculation
from converging to a solution with unphysically low
energy\cite{ding_embedded_2017,prentice_combining_2020}. This
procedure involves forcing the off-diagonal blocks of the overlap
matrix, i.e., $S^{AB}$ and $S^{BA}$, to be zero, by applying a
transformation to the environmental basis functions to ensure they are
orthogonal to the active region's basis functions. For more details on
block orthogonalisation, see Refs.\ \citenum{ding_embedded_2017} and
\citenum{prentice_combining_2020}. This block orthogonalisation
procedure is applied throughout this work, and its effect on accuracy
is discussed in Section \ref{subsec:Dimer}.

\subsection{Time-dependent embedded mean-field theory} \label{subsec:TDEMFT}

Because EMFT is a mean-field theory, like DFT, TD-EMFT can be derived
using a very similar process to that of standard linear-response
TDDFT\cite{ding_linear-response_2017}, which is briefly outlined in
Section S2 of the Supporting Information. The key quantities here are
the exchange-correlation kernel $f_{xc}(\mathbf{r},\mathbf{r}')$ and
the coupling matrix
$Q_{cv,c'v'}$\cite{zuehlsdorff_linear-scaling_2013}, defined as
\begin{equation}
  f_{xc}(\mathbf{r},\mathbf{r}') = \frac{\delta^2 E_{xc} [\rho]}{\delta\rho(\mathbf{r}) \delta\rho(\mathbf{r}')} ~, \label{eq:XCKernelMainText}
\end{equation}
\begin{equation}
  Q_{cv,c'v'} = \iint d^3\mathbf{r} \, d^3\mathbf{r}' \, \psi^*_c(\mathbf{r}) \psi^*_v(\mathbf{r}) \left[ \frac{1}{|\mathbf{r}-\mathbf{r}'|} + f_{xc}(\mathbf{r},\mathbf{r}') \right] \psi_{c'}(\mathbf{r}') \psi_{v'}(\mathbf{r}') ~. \label{eq:CouplingMatDefMainText}
\end{equation}
$E_{xc}$ is the exchange-correlation energy, $\rho(\mathbf{r})$ is the
electronic density, and $\psi_v(\mathbf{r})$ and $\psi_c(\mathbf{r})$
represent valence and conduction Kohn-Sham states, respectively. The
Tamm-Dancoff approximation (TDA)\cite{hirata_time-dependent_1999} (see
Section S2 of the Supporting Information for more details) is also
used throughout this work, which makes calculations substantially more
computationally efficient. Using the TDA can result in some errors in
oscillator strengths relative to solving the full TDDFT problem, but
typically produces reliable excitation
frequencies\cite{zuehlsdorff_linear-scaling_2015}, which are the main
properties of interest in this work.

In order to modify the standard TDDFT procedure for EMFT with
DFT-in-DFT embedding, Eq.\ \eqref{eq:EMFTEnergy} implies that only
changes to $f_{xc}$ need to be considered, as the only thing that
changes between the different levels of theory is $E_{xc}$. Within
EMFT, $E_{xc}^{\text{EMFT}}[\rho] = E_{xc}^{\text{low}}[\rho] + \left(
E_{xc}^{\text{high}} [\rho_{AA}] - E_{xc}^{\text{low}} [\rho_{AA}]
\right)$. If this is substituted into
Eq.\ \eqref{eq:XCKernelMainText}, the result is
\begin{align}
  f_{xc}(\mathbf{r},\mathbf{r}') &= f^{\text{low}}_{xc}(\mathbf{r},\mathbf{r}') + f^{\text{high},AA}_{xc}(\mathbf{r},\mathbf{r}') - f^{\text{low},AA}_{xc}(\mathbf{r},\mathbf{r}') \label{eq:EMFTXCKernel} \\
  &= \frac{\delta^2 E^{\text{low}}_{xc} [\rho]}{\delta\rho(\mathbf{r}) \delta\rho(\mathbf{r}')} + \frac{\delta^2 E^{\text{high}}_{xc} [\rho_{AA}]}{\delta\rho_{AA}(\mathbf{r}) \delta\rho_{AA}(\mathbf{r}')} - \frac{\delta^2 E^{\text{low}}_{xc} [\rho_{AA}]}{\delta\rho_{AA}(\mathbf{r}) \delta\rho_{AA}(\mathbf{r}')} ~. \nonumber 
\end{align}
If this is followed through to the expression for the coupling matrix
$Q$ in Eq.\ \eqref{eq:CouplingMatDefMainText}, $Q$ now
becomes\cite{ding_linear-response_2017}
\begin{align}
  Q_{cv,c'v'} &= \iint d^3\mathbf{r} \, d^3\mathbf{r}' \, \psi^*_c(\mathbf{r}) \psi^*_v(\mathbf{r}) \left[ \frac{1}{|\mathbf{r}-\mathbf{r}'|} + f^{\text{low}}_{xc}(\mathbf{r},\mathbf{r}') \right] \psi_{c'}(\mathbf{r}') \psi_{v'}(\mathbf{r}') \label{eq:EMFTCoupling} \\
  &+ \iint d^3\mathbf{r} \, d^3\mathbf{r}' \, \psi^{A*}_c(\mathbf{r}) \psi^{A*}_v(\mathbf{r}) \left( f^{\text{high},AA}_{xc}(\mathbf{r},\mathbf{r}') - f^{\text{low},AA}_{xc}(\mathbf{r},\mathbf{r}') \right) \psi^A_{c'}(\mathbf{r}') \psi^A_{v'}(\mathbf{r}') ~, \nonumber
\end{align}
where $\psi^A$ represents the projection of a Kohn-Sham eigenstate
onto the basis functions in region $A$ alone. The TDDFT calculation
can now proceed as usual, but using the EMFT result for $Q$
(Eq.\ \eqref{eq:EMFTCoupling}) instead of the standard result
(Eq.\ \eqref{eq:CouplingMatDefMainText}).

Similarly to ground-state DFT, to perform a TDDFT calculation with a
hybrid functional, a fraction of exact exchange must be added to the
coupling matrix $Q$. This contribution, $Q^{\text{EX}}$, can be written
as (in bra-ket
notation)\cite{zuehlsdorff_linear-scaling_2013,fornace_embedded_2015,dziedzic_linear-scaling_2013}
\begin{equation}
  Q^{\text{EX}}_{cv,c'v'} = - 2 \lambda_{\text{EX}} \sum_{\alpha\beta\gamma\delta} \left( \alpha \delta | \beta \gamma \right)
  \braket{\psi_c|\phi^\alpha} \braket{\phi^\beta|\psi_{c'}} \braket{\psi_{v'}|\phi^\delta} \braket{\phi^\gamma|\psi_v} ~. \label{eq:TDDFTHFX}
\end{equation}
As above, $\phi$ are the basis functions, and $\psi_v$ and $\psi_c$
are the valence and conduction Kohn-Sham states respectively. $\left(
\alpha \delta | \beta \gamma \right)$ is an electron repulsion
integral (see
Ref.\ \citenum{dziedzic_linear-scaling_2013}). $\lambda_{\text{EX}}$
corresponds to the fraction of exact exchange included by the hybrid
functional used. In a TD-EMFT calculation with a hybrid functional as
the higher level of theory, this contribution must also be included,
but restricted, as before, to only include exchange within the active
region:
\begin{equation}
  Q^{\text{EX}}_{cv,c'v'} = - 2 \lambda_{\text{EX}}  \sum_{\alpha\beta\gamma\delta\in A} \left( \alpha \delta | \beta \gamma \right)
  \braket{\psi_c|\phi^\alpha} \braket{\phi^\beta|\psi_{c'}} \braket{\psi_{v'}|\phi^\delta} \braket{\phi^\gamma|\psi_v} ~. \label{eq:TDEMFTHFX}
\end{equation}
By replacing the contribution given by Eq.\ \eqref{eq:TDDFTHFX} with
that in Eq.\ \eqref{eq:TDEMFTHFX}, established methods to solve the
hybrid TDDFT problem can be used for TD-EMFT calculations.

\subsection{TD-EMFT in \textsc{onetep}} \label{subsec:ONETEP}

In common with many other DFT codes, \textsc{onetep} uses a set of
atom-centred basis functions to describe the
system\cite{prentice_onetep_2020}. What makes \textsc{onetep}
different, however, is that these basis functions are not fixed --
they are individually optimised to reflect the local environment of
the atom on which they are situated, by optimising the energy with
respect to both the density kernel and the form of the basis functions
themselves\cite{prentice_onetep_2020}. Doing this allows for the basis
set to be minimal in size, whilst still maintaining excellent
accuracy. These basis functions, known as non-orthogonal generalised
Wannier functions (NGWFs), are not required to be orthogonal to each
other, and are strictly localised, meaning they are defined to be zero
beyond a certain radius from the atom they are centred on. This
localisation means that matrices, such as the Hamiltonian, are sparse,
and therefore sparse matrix algebra can be used to improve the
efficiency of the calculation. To allow for optimisation, the NGWFs
are defined on an underlying basis of psinc functions. The number of
functions in this underlying basis is controlled by a cutoff energy,
in an analogous way to the same quantity in plane-wave basis sets.

The details of the implementation of ground-state EMFT in
\textsc{onetep} are presented in detail in
Ref.\ \citenum{prentice_combining_2020}, but here, one particular
point of importance should be restated. As described in
Ref.\ \citenum{prentice_combining_2020}, although it is possible to
optimise the NGWFs within an EMFT framework, the introduction of block
orthogonalisation (see Section \ref{subsec:EMFT}), significantly
affects this optimisation. Block orthogonalisation effectively adds a
new term to the gradient used to optimise the NGWFs; this new term
competes with the other terms, leading to the optimisation
stalling. To avoid this, the NGWFs for the whole system are optimised
at the lower level of theory (without imposing block
orthogonalisation), before fixing the NGWFs, block orthogonalising
them, and optimising the density kernel with EMFT. Although this means
that the NGWFs are not completely optimised at the EMFT level, this
gives an error in the total energy of less than 1\%, which still
provides excellent accuracy\cite{prentice_combining_2020}. The
relative cost of this final optimisation of the density kernel using
EMFT varies depending on system size and parallelisation, but for the
explicitly solvated phenolphthalein system discussed in Section
\ref{subsec:PhPh} and treated with PBE0-in-PBE EMFT, this step takes
roughly twice as long as an optimisation at the lower level of theory.

In a \textsc{onetep} ground-state energy calculation, the NGWFs are
optimised to describe the occupied, or valence, Kohn-Sham states. This
means there is no guarantee that these NGWFs will describe the
unoccupied, or conduction, states, and indeed this is often the
case\cite{skylaris_using_2005,ratcliff_calculating_2011,prentice_onetep_2020}. However,
describing the conduction states well, or at least a subset of them,
is vital for performing accurate calculations of excited-state
properties. To remedy this, when such calculations are required, a new
set of NGWFs is created to describe the conduction states. These
conduction NGWFs are optimised to describe a given number of the
lowest-lying conduction states, by projecting the valence states out
of the Hamiltonian. The original set of `valence' NGWFs and the new
conduction NGWFs are then combined into a joint NGWF basis set that
can describe both valence and conduction
states\cite{ratcliff_calculating_2011,ratcliff_ab_2013}. The same
procedure is followed in a TD-EMFT calculation, simply projecting the
valence states out of the EMFT Hamiltonian. As with the valence NGWFs,
block orthogonalisation is applied, implying that the conduction NGWFs
are optimised at the lower level of theory only, before fixing them
and optimising the conduction density kernel with EMFT.

Once a set of basis functions that can be used to correctly describe
both valence and conduction Kohn-Sham states has been obtained, TDDFT
calculations can be performed. TDDFT calculations in \textsc{onetep}
follow the algorithm laid out in
Ref.\ \citenum{zuehlsdorff_linear-scaling_2013}, which is briefly
outlined in Section S3 of the Supporting Information. To modify this
algorithm for TD-EMFT, as in Section \ref{subsec:TDEMFT} the usual
expression for $f_{xc}$ is replaced with the EMFT expression for this
quantity, shown in Equation \eqref{eq:EMFTXCKernel}.

Although it is not used in the results presented in this work, a
feature of the TDDFT implementation in \textsc{onetep} relevant to
TD-EMFT should still be emphasised. Because of the different levels of
theory used to treat the active region and the environment in TD-EMFT,
spurious excitations involving charge transfer between the two regions
can become possible, particularly if the introduction of EMFT results
in energy levels associated with different regions to swap their
ordering. However, by truncating the response density kernel
appropriately within \textsc{onetep}, it is possible to exclude
particular types of excitations from the calculation -- for example,
non-physical low-energy charge-transfer states that are a known issue
with semi-local TDDFT\cite{zuehlsdorff_solvent_2016}. In particular,
the excitations can be forced to be localised on a specific set of
atoms, by setting to zero any element of the response density kernel
that involves a basis function not associated with these atoms. This
would allow for spurious charge transfer between the regions in
TD-EMFT, if present, to be eliminated, by localising the excitations
on the active region alone. However, the systems tested in this work
do not exhibit such unphysical charge transfer excitations, and
therefore no response kernel truncation is applied.

\subsection{Combining (TD)-EMFT and implicit solvation} \label{subsec:ImpSol}

An implicit solvation model is included within \textsc{onetep}, using
a minimal parameter solvent model based on the model of Fattebert and
Gygi\cite{fattebert_density_2002,fattebert_first-principles_2003} and
extended by Scherlis \textit{et
  al}\cite{scherlis_unified_2006,dziedzic_minimal_2011,dziedzic_large-scale_2013}. This
model allows the solvent environment of the system under study to be
described classically, as a polarisable dielectric medium. A cavity is
defined around the system; at the edge of this cavity, there is a
smooth transition in the value of the dielectric permittivity from the
vacuum to the appropriate value for the solvent in
question\cite{fattebert_density_2002}. The size and shape of the
cavity is determined by the electronic density -- typically a
preliminary ground-state calculation is performed with the system in
vacuum, before defining and fixing the cavity at the size and shape
implied by the density calculated in
vacuum\cite{dziedzic_minimal_2011}. The polarisation induced in the
solvent medium by the distribution of charge in the system can be
calculated, and this in turn induces a new potential that is included
in the Hamiltonian when optimising the density kernel and NGWFs. This
means that the electrostatic interaction between the solvent and the
system can be included self-consistently when determining the
ground-state energy, valence NGWFs, and conduction NGWFs, as well as
the excitation spectrum.

Using the implicit solvation model in \textsc{onetep} currently
requires the use of open boundary conditions (OBCs), rather than the
periodic boundary conditions (PBCs) typical in the rest of the
code. Under these conditions, \textsc{onetep} uses the \textsc{dl\_mg}
library to calculate the total electrostatic potential by solving the
generalised Poisson equation\cite{womack_dl_mg_2018}. \textsc{dl\_mg}
is a multi-grid method. This solver also allows for the treatment of
OBC calculations in vacuum within \textsc{onetep}, as such
calculations just correspond to implicit solvent calculations with a
solvent with a permittivity of 1.

For a given electronic density and cavity, the potential induced by
the polarisation of the solvent is independent of the functional used
in the rest of the calculation. This means that the use of (TD-)EMFT
does not affect the implicit solvation model directly, only indirectly
by producing a different electronic density to the regular
Hamiltonian. The main subtlety lies in how the cavity is defined. The
preliminary calculation in vacuum used to determine this can be
performed either at the lower level of theory, or using EMFT. Both
methods will give the same set of NGWFs (as NGWFs are optimised at the
lower level of theory, as previously mentioned), but different density
kernels. These two kernels will give slightly different cavities, and
potentially therefore different results. This difference is explored
in Sections \ref{subsec:Dimer} and \ref{subsec:PhPh}.

\section{Results} \label{sec:Examples}

To test the implementation of TD-EMFT within linear-scaling DFT
(specifically the code \textsc{onetep}), I have applied it to several
different systems. In this section, each of these systems is described
in turn, and the results of TD-EMFT calculations are reported, in
order to validate and demonstrate the capabilities of the
implementation. The \texttt{.cif} files for all structures shown are
provided in the Supporting Information, converted using
\textsc{c2x}\cite{rutter_c2x_2018}. All spectra are broadened using
lifetime broadening -- \textsc{ONETEP} calculates the lifetime of each
excitation, and applies a Lorentzian broadening function to each
excitation energy with a width corresponding to the appropriate
lifetime.

\subsection{Water-nitrogen dimer} \label{subsec:Dimer}

\begin{figure}
  \centering
  \includegraphics[trim={6cm 5.8cm 6cm 6cm},clip,width=0.5\textwidth]{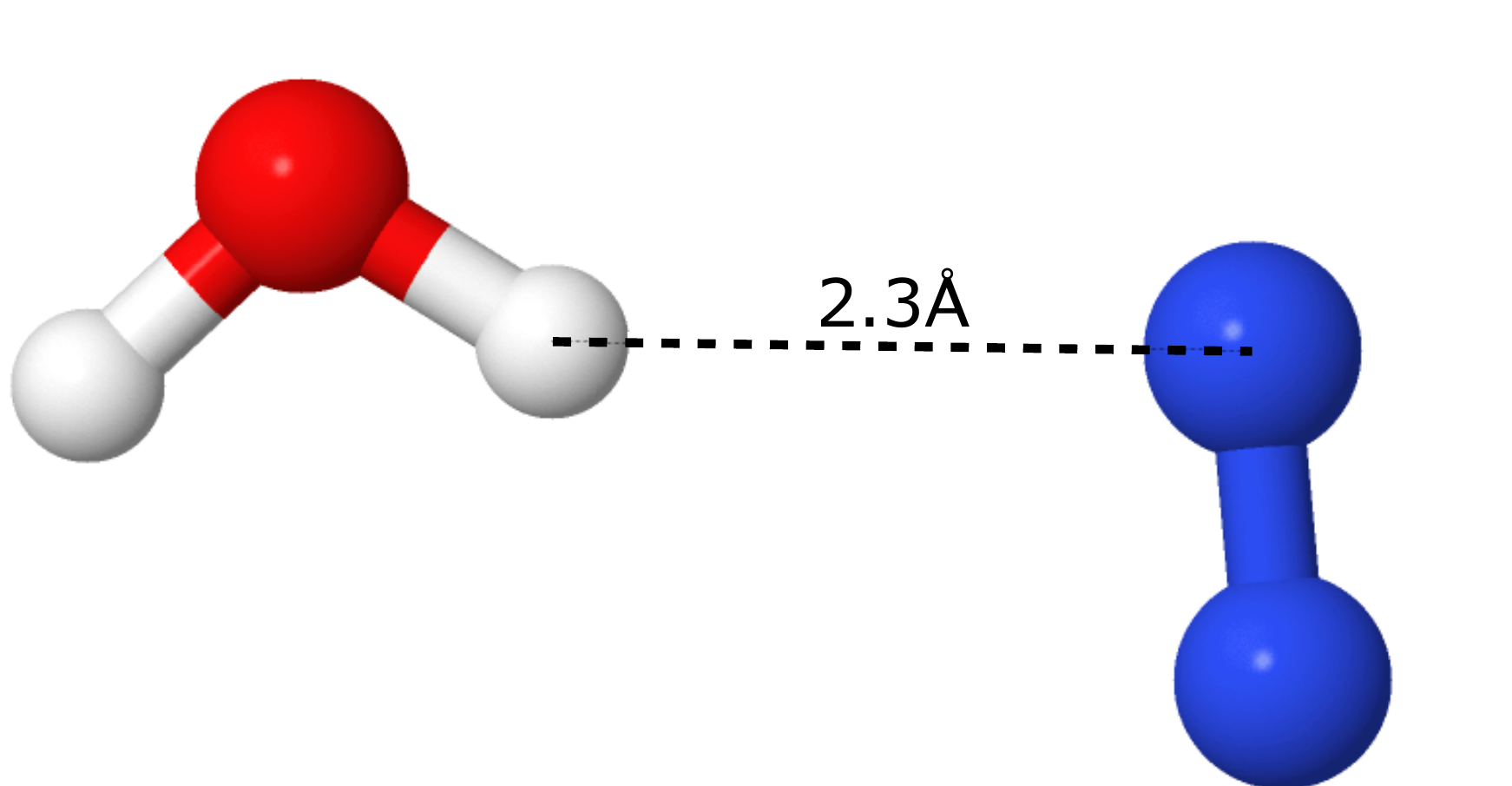}
  \caption{The structure of the water-nitrogen dimer used in this
    work. O, H, and N atoms are red, white, and blue respectively. The
    shortest distance between the molecules is 2.3~\r{A}, as labelled
    on the figure. Figure produced using Jmol\cite{hanson_jmol_2010}.}
  \label{fig:WaterN2DimerStructure}
\end{figure}

I first tested TD-EMFT as implemented within linear-scaling DFT on a
very small and simple system -- a dimer composed of a water molecule
and a nitrogen molecule, separated by a distance of 2.3~\r{A}. This
structure is shown in Fig.\ \ref{fig:WaterN2DimerStructure}. I chose a
dimer containing two different molecules rather than a water dimer to
enable examination of any change in behaviour when I changed which
molecule is designated as the active region.

I examined the dimer both in vacuum and in implicit solvent, where the
parameters of the implicit solvent are those appropriate for water
near room temperature (permittivity $\epsilon_r=78.54$, surface
tension $\gamma=0.07415$~N~m$^{-1}$). This allowed testing of the
TD-EMFT implementation both with and without implicit solvent. I also
looked at the effect of defining the implicit solvent cavity using the
kernel optimised with EMFT (referred to as the EMFT cavity), or using
the kernel optimised at the lower level of theory only (the non-EMFT
cavity), as discussed in Section \ref{subsec:ImpSol}.

The lower level of theory was chosen to be the local density
approximation
(LDA)\cite{kohn_self-consistent_1965,perdew_self-interaction_1981},
whilst the higher level of theory was chosen to be the widely-used
hybrid functional
B3LYP\cite{becke_density-functional_1993}. Norm-conserving
pseudopotentials distributed with \textsc{onetep} were used for all
three species. A cut-off energy of $850$~eV was used, and NGWF radii
of $11$~bohr were used for all species. 4 NGWFs were associated with
each of the O and N atoms, and 1 with the H atoms. The dimer was
centered in a large cubic cell, with side lengths of $75$~bohr. The
vacuum calculations were performed under PBCs in this cell, whilst the
implicit solvent calculations were performed under OBCs. In the vacuum
calculations, the large size of the cell eliminates interaction
between the dimer and its periodic images, meaning that they are
directly comparable to the implicit solvent calculations. OBC
calculations within ONETEP must make use of the DL\_MG multigrid
solver, which reduces their computational efficiency somewhat, so
large-cell PBC calculations are preferred where possible.

\begin{figure*}
  \centering
  \subcaptionbox{Vacuum}{
    \includegraphics[width=0.45\textwidth]{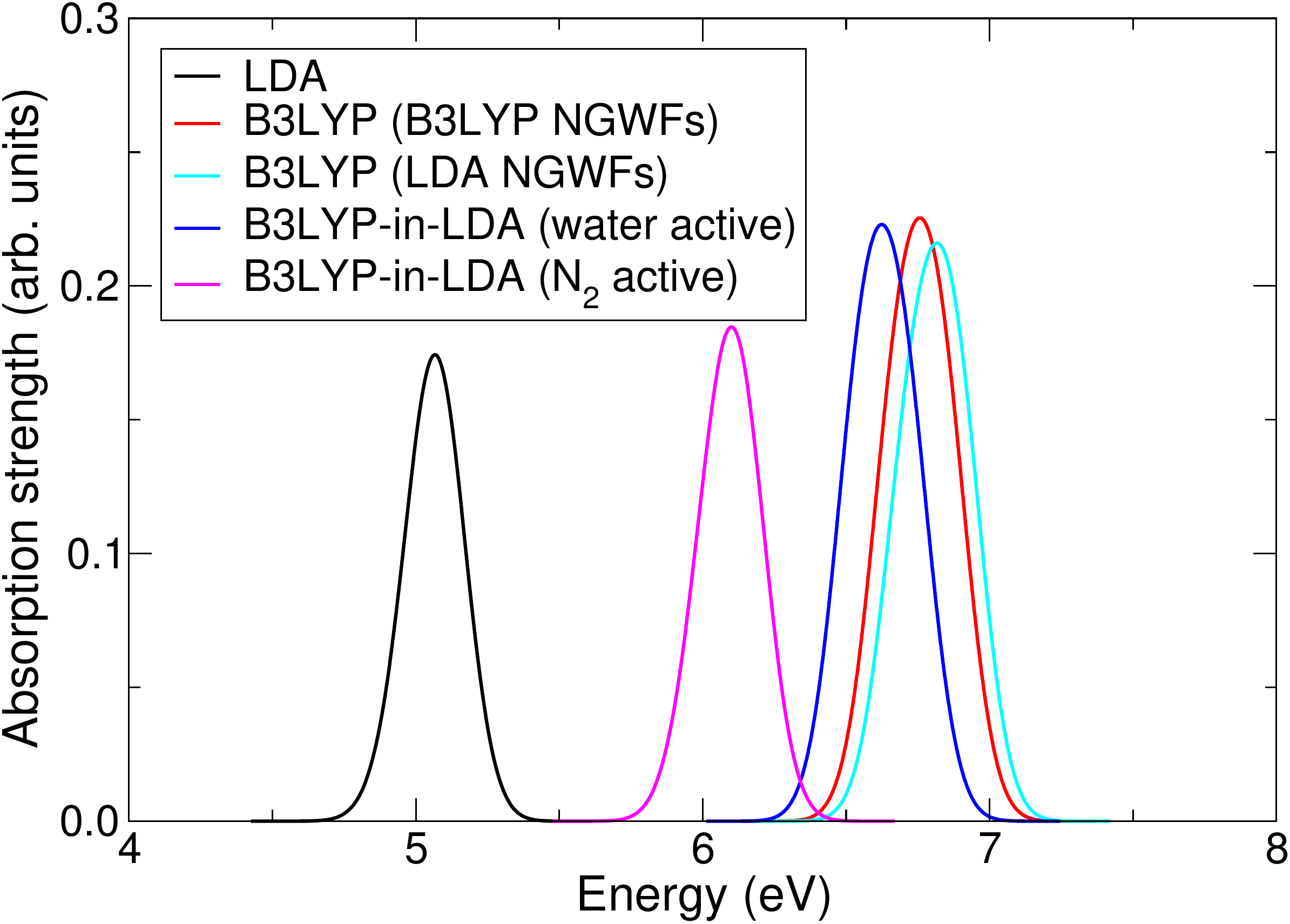}
  }
  ~
  \subcaptionbox{Implicit solvent}{
    \includegraphics[width=0.45\textwidth]{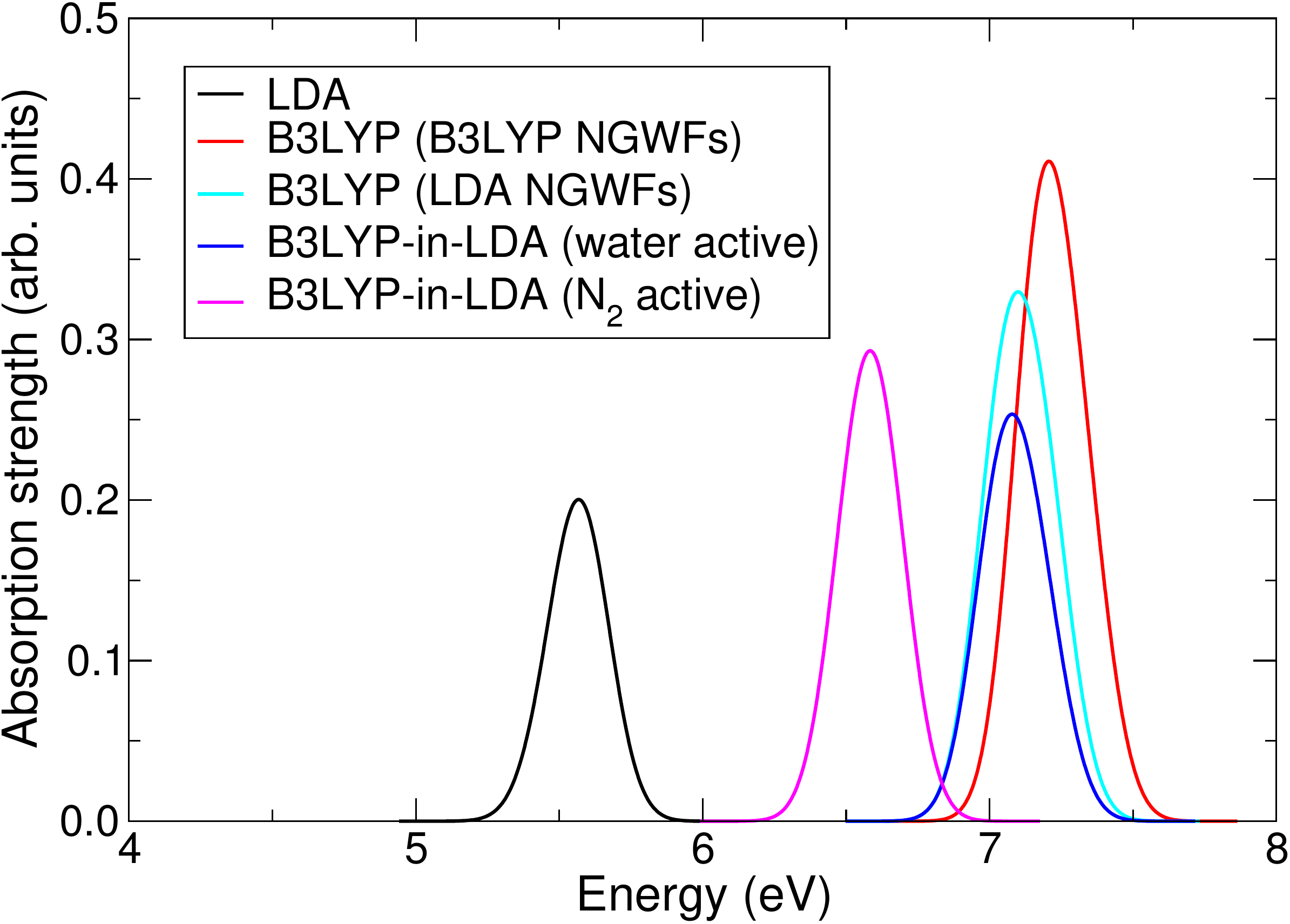}
  }
  \caption{Low-energy absorption spectra of the water-nitrogen dimer, calculated
    at various levels of theory: full system LDA (black), full system
    B3LYP with LDA-optimised NGWFs (cyan), full system B3LYP with
    B3LYP-optimised NGWFs (red), B3LYP-in-LDA with water as the active
    region (blue), and B3LYP-in-LDA with nitrogen as the active region
    (magenta). a) shows the results calculated in vacuum, and b) shows
    the results calculated in implicit solvent (water), using the
    non-EMFT cavity.}
  \label{fig:WaterN2DimerSpectra}
\end{figure*}

\begin{figure}
  \centering
  \includegraphics[width=0.5\textwidth]{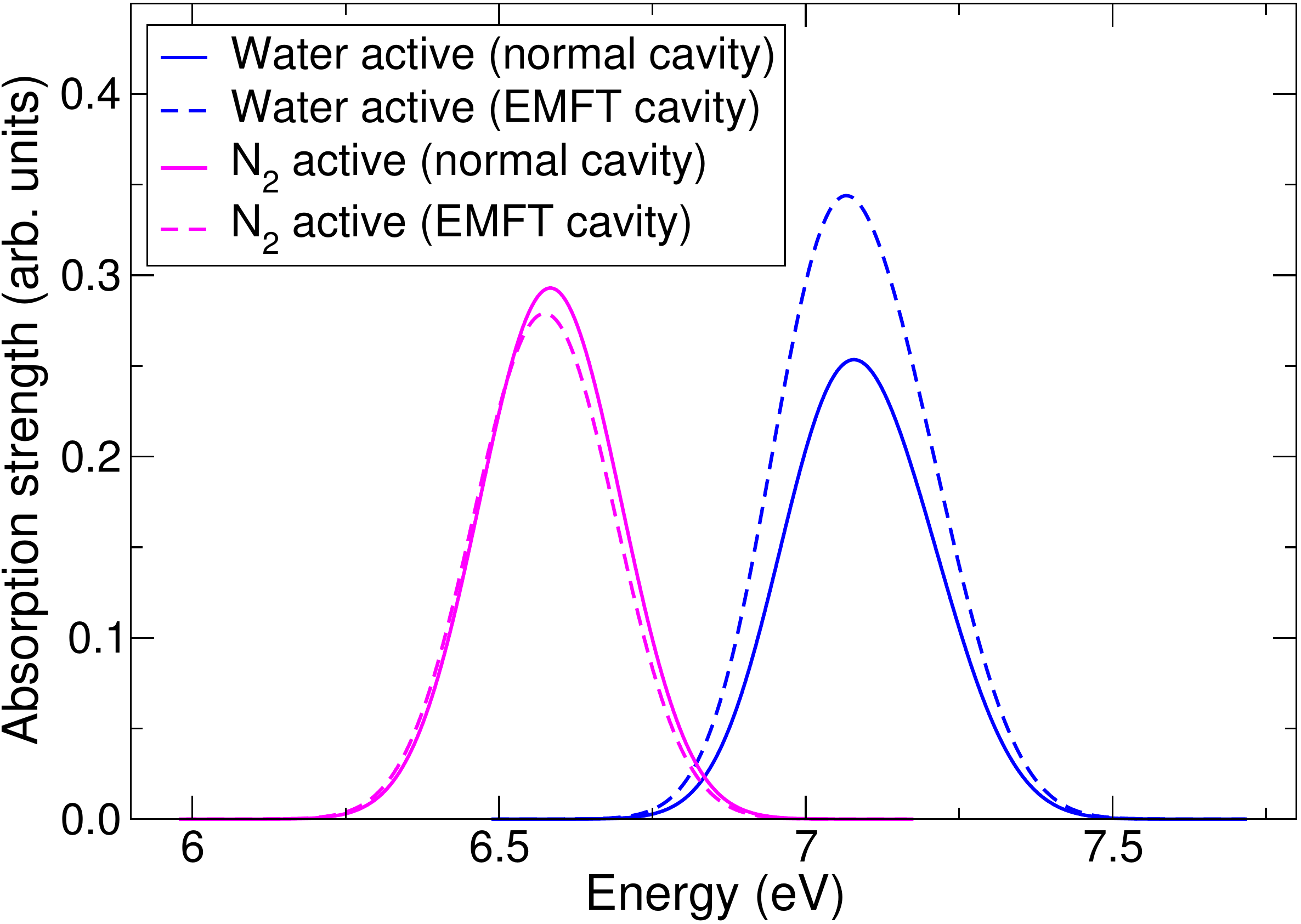}
  \caption{Low-energy absorption spectra of the water-nitrogen dimer, calculated
    using B3LYP-in-LDA TD-EMFT in implicit solvent, using either the
    non-EMFT (solid lines) or EMFT cavity (dashed lines). The results
    for both B3LYP-in-LDA with water as the active region (blue
    lines), and B3LYP-in-LDA with nitrogen as the active region
    (magenta lines), are presented.}
  \label{fig:WaterN2DimerCavitySpectra}
\end{figure}

Most of the results presented here are focused on the lowest energy
reasonably bright excitation of the water-nitrogen dimer, which ranges
between $5$ and $7$~eV in vacuum depending on the method used. This
excitation is not the strongest in the spectrum of this system --
there is another brighter excitation that ranges between $6.5$ and
$7.5$~eV in vacuum. However, focusing on the lower energy excitation
allows for a more thorough test of TD-EMFT, as the difference between
LDA and B3LYP is very pronounced for this excitation. This difference
is not as large for the higher energy excitation, although the same
conclusions can be drawn from both. The higher energy excitation is
discussed at the end of the present section, as well as in Section S4
in the Supporting Information.

The effect of the block orthogonalisation (BO) protocol discussed in
Section \ref{subsec:EMFT} can be identified by comparing a standard
LDA TDDFT calculation, and an LDA-in-LDA TD-EMFT calculation. An
LDA-in-LDA TD-EMFT calculation treats all parts of the system at the
same level of theory (LDA), but does so using the machinery of EMFT,
including, importantly, BO -- this means that any difference between
the calculations can be ascribed to the presence of BO. A comparison
between these calculations for the water-nitrogen dimer shows there is
excellent agreement. The difference in ground state energies is $1.3$
and $2.4$~meV in vacuum and solvent respectively, well within
acceptable limits. This is also the case for the low-energy excitation
-- the difference in excitation energy is $2.1$ and $3.3$~meV in
vacuum and solvent respectively. This demonstrates that BO does not
significantly affect the accuracy of the results. The conduction NGWFs
seem to be more sensitive to the presence of BO -- if the `ground
state' energy of the projected Hamiltonian used to optimise the
conduction NGWFs\cite{ratcliff_calculating_2011} is compared, the
difference is $52$ and $48$~meV in vacuum and solvent
respectively. However, this level of agreement is still more than
sufficient to obtain matching solutions to the TDDFT problem, as
already seen.

Fig.\ \ref{fig:WaterN2DimerSpectra} presents the calculated low-energy
absorption spectra -- Fig.\ \ref{fig:WaterN2DimerSpectra}a shows the
results in vacuum, whilst Fig.\ \ref{fig:WaterN2DimerSpectra}b shows
them in implicit solvent, using the non-EMFT cavity. The low-energy
absorption spectrum is calculated using LDA for the whole system
(black line in figures), B3LYP for the whole system (red/cyan), and
B3LYP-in-LDA EMFT, with either the water or nitrogen molecule acting
as the active region (blue and magenta respectively). When the whole
system is treated with B3LYP, two sets of results are presented -- one
using NGWFs optimised at the B3LYP level (red), as in a normal
\textsc{onetep} calculation, and one using NGWFs optimised at the LDA
level, as in a EMFT \textsc{onetep} calculation (cyan).

As expected, LDA produces a significantly lower excitation energy than
B3LYP in all cases -- this is precisely the discrepancy TD-EMFT aims
to correct. It can also be seen that the B3LYP calculations performed with
LDA- and B3LYP-optimised NGWFs agree well; the excitation energy
calculated with LDA-optimised NGWFs is within $0.05$~eV of that
obtained in the pure B3LYP calculation in vacuum, and within $0.13$~eV
in implicit solvent. This demonstrates that the error introduced by
using NGWFs optimised at the lower level of theory does not
significantly affect the accuracy of the calculation, especially when
compared to the difference between the lower and higher levels of
theory, validating this approximation within the TD-EMFT calculations.

\begin{figure*}
  \centering
  \subcaptionbox{LDA}{
    \includegraphics[trim={18cm 7cm 18cm 8cm},clip,width=0.45\textwidth]{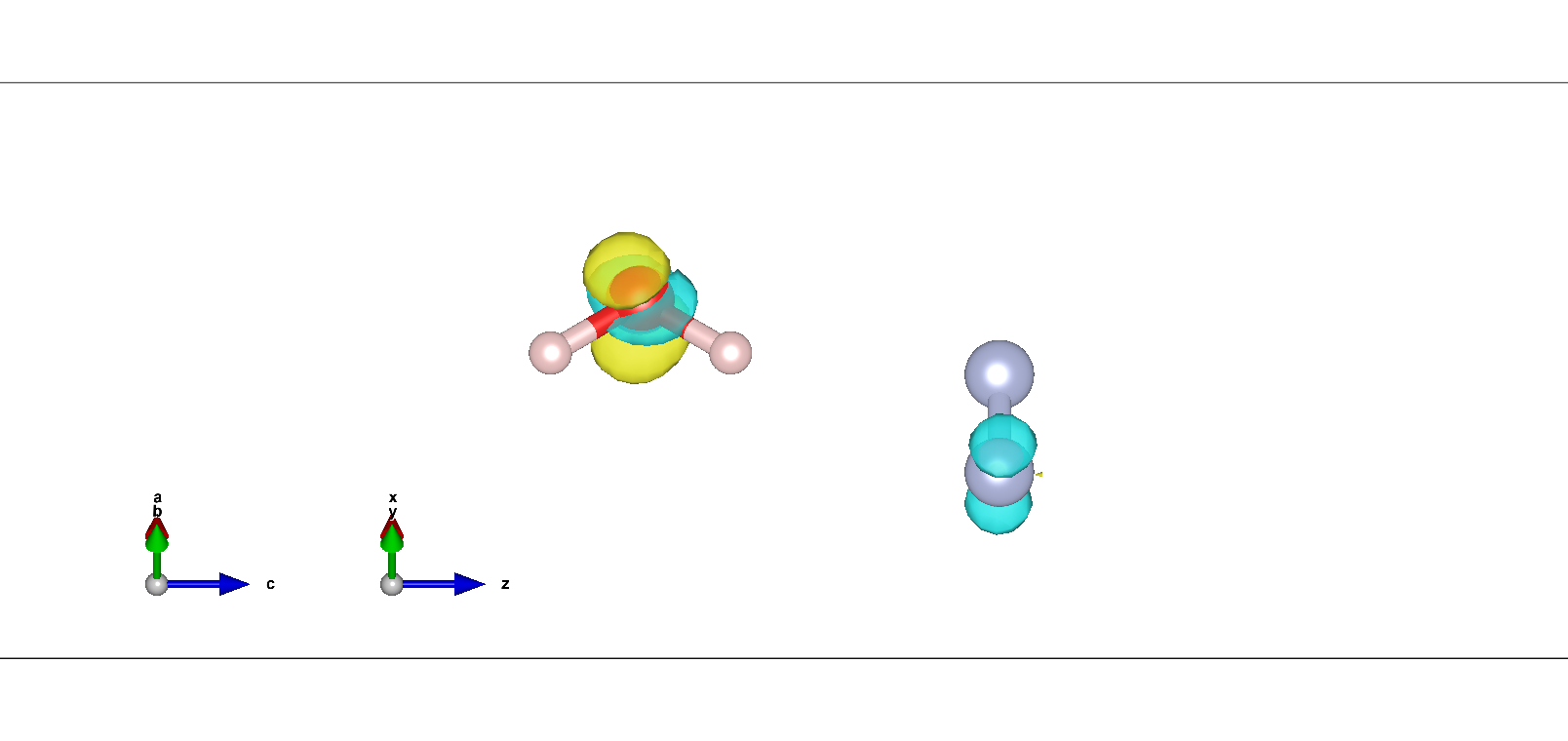}
  }
  ~
  \subcaptionbox{B3LYP}{
    \includegraphics[trim={18cm 7cm 18cm 8cm},clip,width=0.45\textwidth]{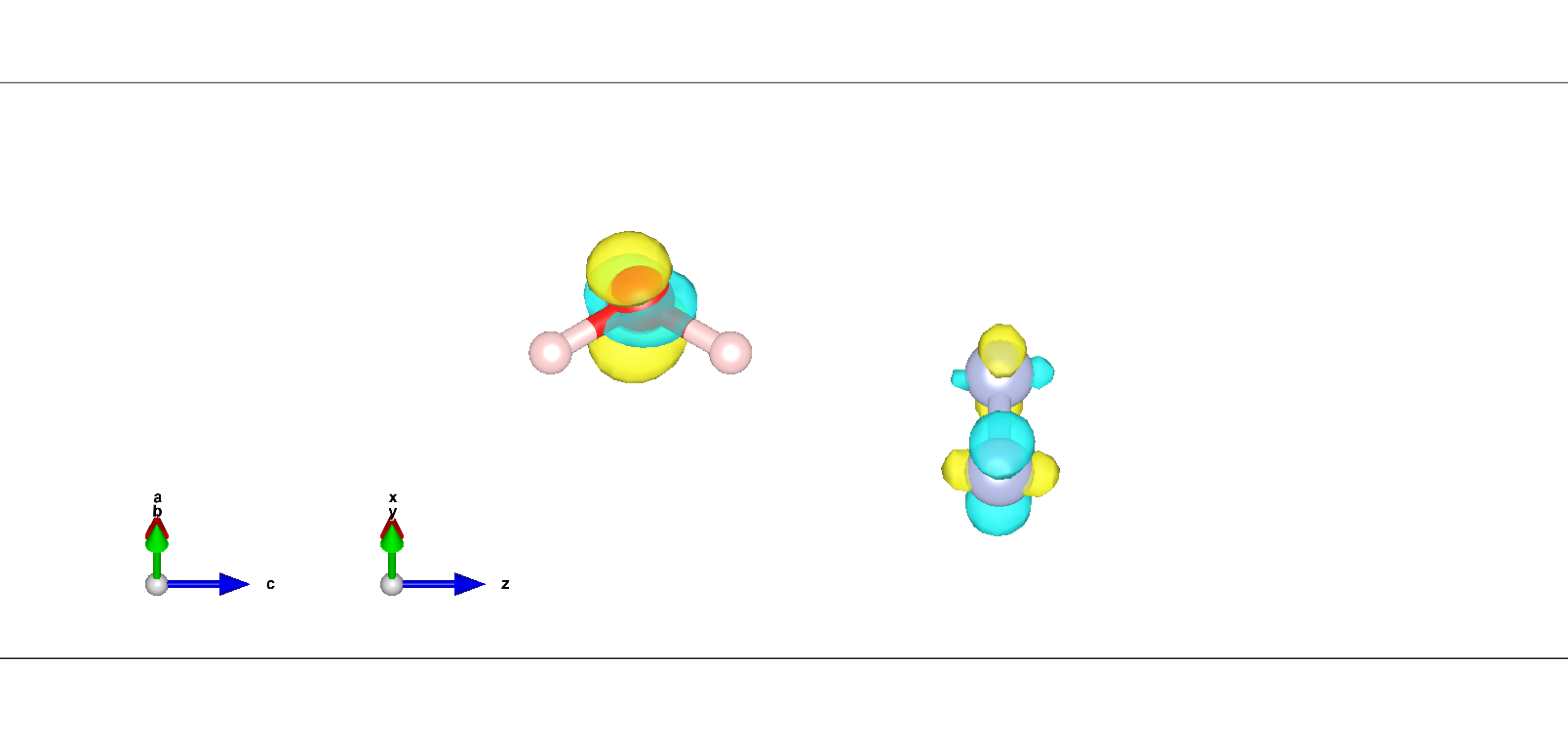}
  }
  ~
  \subcaptionbox{B3LYP-in-LDA (water active)}{
    \includegraphics[trim={18cm 7cm 18cm 8cm},clip,width=0.45\textwidth]{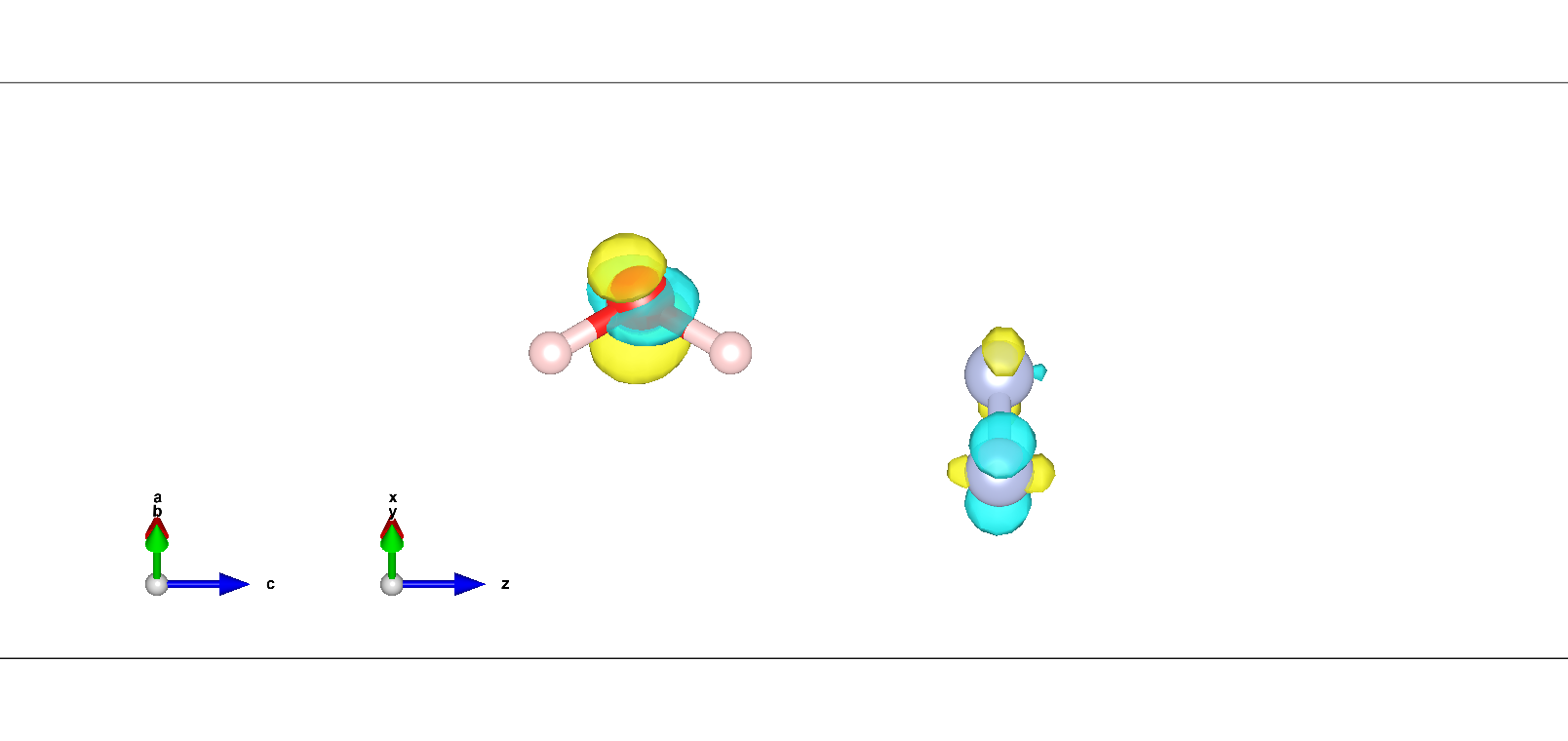}
  }
  ~
  \subcaptionbox{B3LYP-in-LDA (N$_2$ active)}{
    \includegraphics[trim={18cm 7cm 18cm 8cm},clip,width=0.45\textwidth]{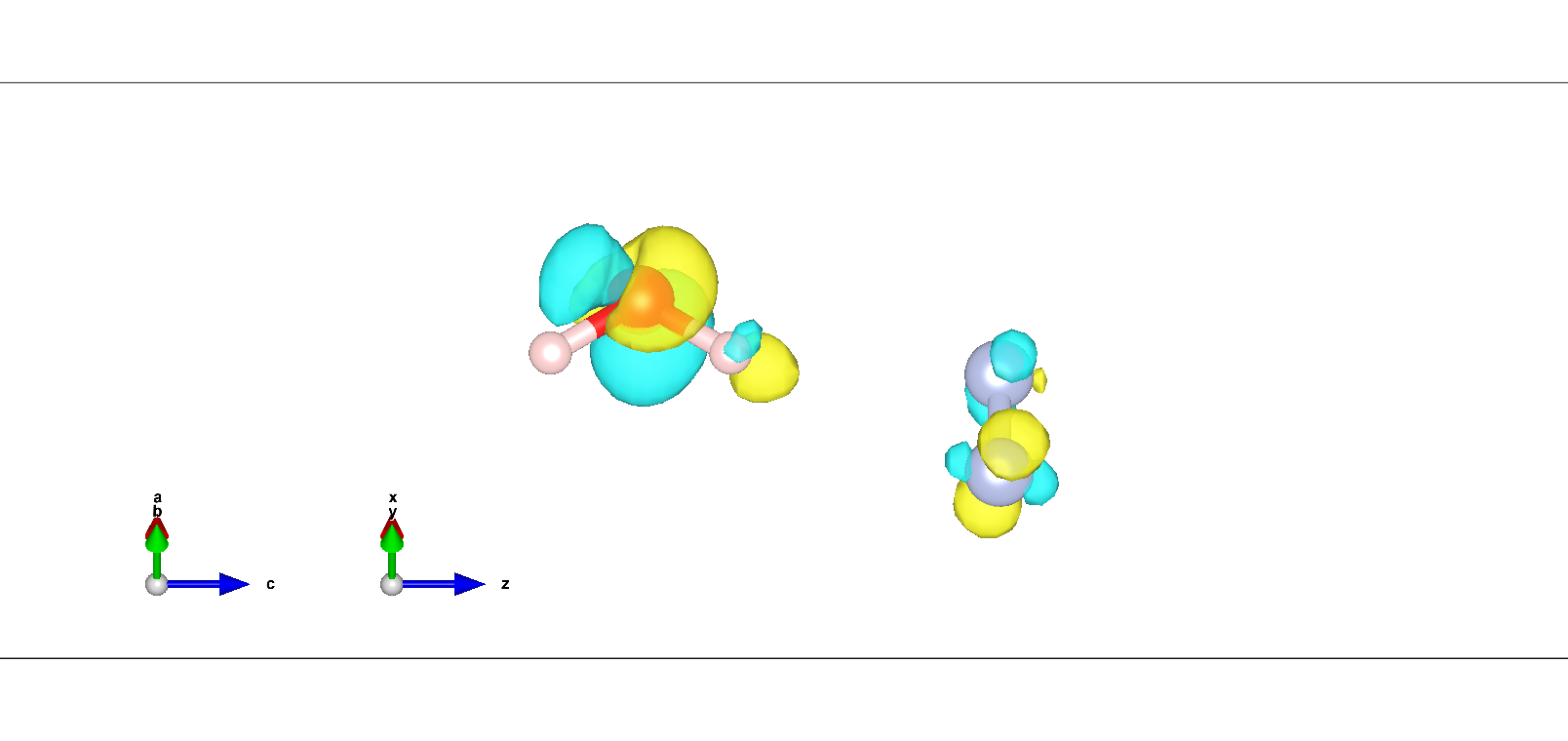}
  }
  \caption{Isosurfaces of the calculated response density for the
    excitation of the water-nitrogen dimer seen in
    Fig.\ \ref{fig:WaterN2DimerSpectra}b, calculated at various levels
    of theory: full system LDA, full system B3LYP (with
    B3LYP-optimised NGWFs), B3LYP-in-LDA with water as the active
    region, and B3LYP-in-LDA with nitrogen as the active region. All
    response densities were computed with implicit solvent, using the
    non-EMFT cavity. The isosurfaces are at $|n|=0.01$~e\,\r{A}$^{-3}$,
    with yellow and blue representing positive and negative response
    densities respectively. O, N, and H atoms are red, blue, and white
    respectively. Figures produced using
    VESTA\cite{momma_vesta_2011}.}
  \label{fig:WaterN2DimerCharacter}
\end{figure*}

However, the most important feature of the spectra shown in
Fig.\ \ref{fig:WaterN2DimerSpectra} is that the B3LYP-in-LDA results
agree well with the full system B3LYP results, if the water molecule
is taken as the active region. If the water molecule is treated with
B3LYP, TD-EMFT calculations give an error compared to the full B3LYP
results of $0.13$~eV in vacuum and $0.15$~eV in implicit solvent --
comparable to the error arising from using LDA-optimised NGWFs. If
instead the nitrogen molecule is taken as the active region, this
error becomes significantly worse, although the resulting excitation
energy is still significantly closer to the B3LYP value than the LDA
value. The reasons for this are discussed in more detail below, with
reference to Fig.\ \ref{fig:WaterN2DimerCharacter}. The oscillator
strength of the excitation also varies a little as the level of theory
is changed, although this is a secondary concern as the oscillator
strength is less reliably calculated under the TDA anyway. Taken
together, these results validate the accuracy of the TD-EMFT method,
and in particular the implementation of it in \textsc{onetep},
\textit{as long as} the active region is chosen wisely.

Examining the effect of implicit solvent on our calculations,
comparing Figs.\ \ref{fig:WaterN2DimerSpectra}a and
\ref{fig:WaterN2DimerSpectra}b shows that that the introduction of
implicit solvent induces a blue shift of roughly $0.5$~eV at every
level of theory, and also has some effect on the oscillator
strengths. The overall accuracy of the TD-EMFT method, however, is not
significantly affected by the presence of implicit solvent,
demonstrating that TD-EMFT and implicit solvent can be used
successfully together. Fig.\ \ref{fig:WaterN2DimerCavitySpectra} shows
that changing whether the cavity is created using the LDA- or
EMFT-optimised density kernel makes very little difference to the
excitation energies, but can change the absorption strengths. This is
likely a symptom of the fact that the active region is at the edge of
the cavity, so the change in kernel will directly affect the shape and
size of the cavity. This is not a particularly likely mode of
operation -- in more realistic systems, such as the phenolphthalein
system in Section \ref{subsec:PhPh}, the active region will be
surrounded by the lower-level environment region, and will therefore
not be close to the edge of the cavity. In such systems, results
obtained with the EMFT and non-EMFT cavities would be expected to be
extremely similar. It is reassuring, however, that even in the case
where the active region does lie at the edge of the cavity, the choice
of cavity does not affect the accuracy of the most important property
-- the calculated excitation energies.

The results shown in Figs.\ \ref{fig:WaterN2DimerSpectra} and
\ref{fig:WaterN2DimerCavitySpectra} can be more clearly understood by
looking more closely at the character of these
excitations. Fig.\ \ref{fig:WaterN2DimerCharacter} shows isosurfaces
of the response density for the excitation calculated at different
levels of theory. It is immediately obvious that in all cases, the
excitation has character on both the water and nitrogen molecules, but
that the response density is higher near the water molecule. This
implies that the excitation is more associated with the water molecule
than the nitrogen molecule. This fits with
Fig.\ \ref{fig:WaterN2DimerSpectra}, where the B3LYP-in-LDA results
are much closer to the full B3LYP results when the water molecule is
the active region. The excitation looks very similar in the LDA,
B3LYP, and B3LYP-in-LDA (active water) calculations
(Figs.\ \ref{fig:WaterN2DimerCharacter}a, b, and c respectively). This
emphasises that, if the most important region is treated at the higher
level of theory, an accurate description of the system can be obtained
with TD-EMFT. However, if the nitrogen molecule is treated as the
active region instead, the excitation changes quite significantly, as
can be seen in Fig.\ \ref{fig:WaterN2DimerCharacter}d. This shows that
describing the wrong part of the system at the higher level of theory
can change the nature of the excitation. Because the excitation does
have some character on the nitrogen molecule, the active nitrogen
TD-EMFT calculation would be expected to correct the LDA excitation
energy to some extent, as can be seen in
Fig.\ \ref{fig:WaterN2DimerSpectra}a, but not to the same extent as
the active water TD-EMFT calculation. This also implies that some of
the error in the B3LYP-in-LDA TD-EMFT results likely comes from the
incorrect treatment of the part of the excitation that is localised on
the molecule treated at the lower level of theory -- this is more of a
problem when the nitrogen molecule is the active region, as previously
discussed.

The discussion above focuses on the low-energy excitations of the
water-nitrogen dimer, as noted previously, but it also applies to the
higher energy bright state found between $6.5$ and $7.5$~eV in
vacuum. Figs.\ S1 and S2 in the Supporting Information give results
for this higher energy excitation, comparable to
Figs.\ \ref{fig:WaterN2DimerSpectra}a and
\ref{fig:WaterN2DimerCharacter} respectively. In this case, the
difference between the energies predicted by LDA and B3LYP (with
B3LYP-optimised NGWFs) is $0.45$~eV, significantly smaller than
before. This means that there is not as much to gain from utilising
TD-EMFT, as LDA describes this excitation much better than the lower
energy excitation treated previously. However, even with this caveat,
B3LYP-in-LDA TD-EMFT is significantly closer to the full B3LYP result,
demonstrating the power of TD-EMFT. When water is the active region,
the error is $0.10$~eV, which is actually less than the error from
using LDA-optimised NGWFs in a B3LYP calculation ($0.15$~eV). Unlike
before, the error in B3LYP-in-LDA calculations with nitrogen as the
active region ($-0.09$~eV) is comparable to calculations with water as
the active region. This is likely partially due to this higher energy
excitation having more response density on the nitrogen molecule, as
seen in Fig.\ S2 in the Supporting Information. Although the gains are
smaller, these data demonstrate the utility of TD-EMFT for the higher
excitation as well. For more discussion on this, see Section S4 in the
Supporting Information.

\subsection{Phenolphthalein in water} \label{subsec:PhPh}

\begin{figure*}
  \centering
  \subcaptionbox{Isolated molecule}{
    \includegraphics[trim={3cm 8cm 4.5cm 8cm},clip,width=0.45\textwidth]{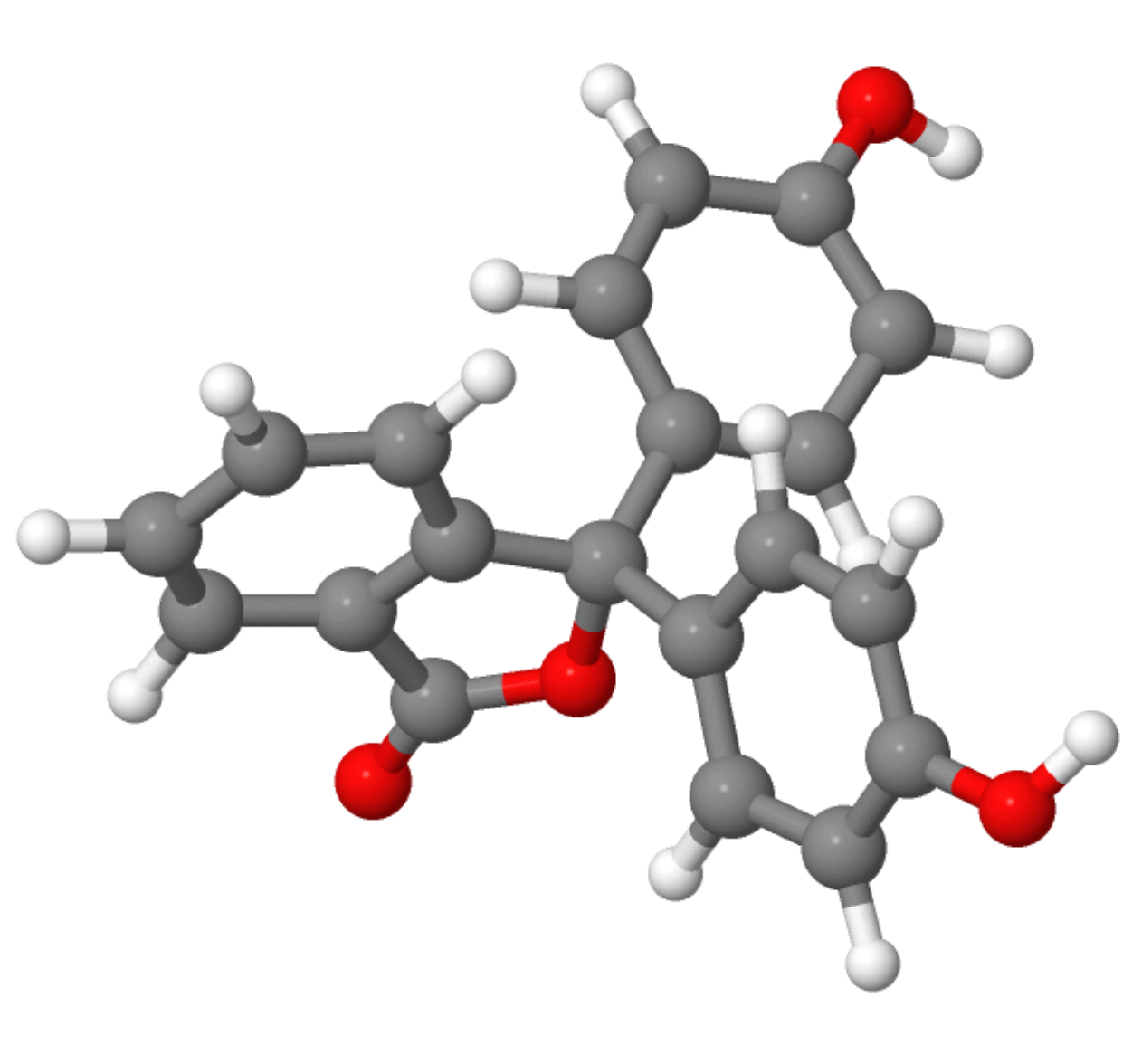}
  }
  ~
  \subcaptionbox{Solvated molecule}{
    \includegraphics[trim={5cm 9cm 5cm 9cm},clip,width=0.45\textwidth]{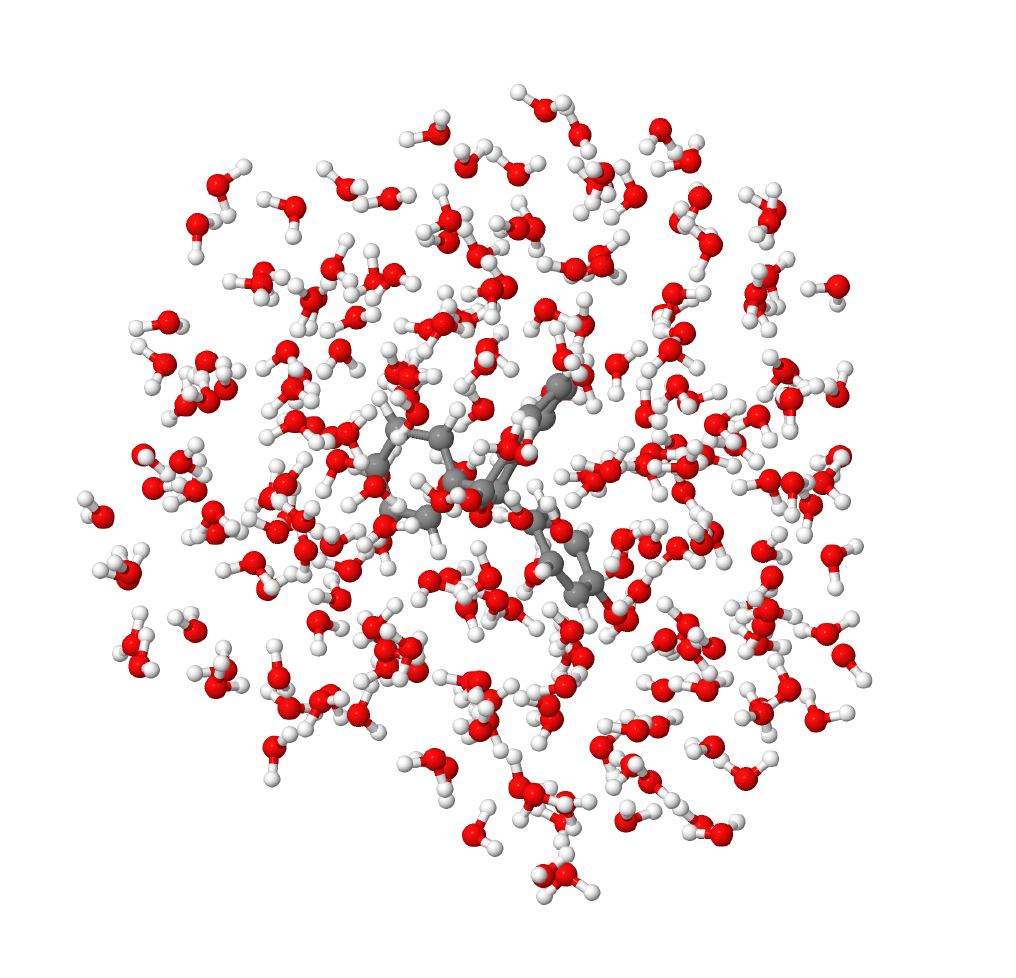}
  }
  \caption{Structures used for the calculations on phenolphthalein in
    this work. a) shows the isolated phenolphthalein molecule, in its
    neutral charge state. b) shows the same molecule explicitly
    solvated with 218 water molecules, as extracted from a classical
    molecular dynamics simulation. C, O, and H atoms are grey, red,
    and white respectively. Figures produced using
    Jmol\cite{hanson_jmol_2010}.}
  \label{fig:PhPhStructures}
\end{figure*}

I next applied linear-scaling TD-EMFT to the case of the molecule
phenolphthalein solvated in water. Phenolphthalein is a well-known pH
indicator. For pH values up to around $8$-$9$, the molecule is in a
charge-neutral configuration that is colourless, but at higher pH, it
donates two protons, becoming doubly negatively charged. This changes
the chemical structure of the molecule and leads to it exhibiting a
fuschia-pink colour\cite{wittke_reactions_1983}. In this work, I
focused on the neutral configuration, which is shown in
Fig.\ \ref{fig:PhPhStructures}a.

Phenolphthalein is typically used as an indicator in solution, usually
water. In order to accurately describe the absorption spectrum of
phenolphthalein, therefore, the effect of the solvent must be
accurately described. This requires not only an excellent quantum
mechanical description of the interaction between the solvent and
solute, but also an accurate description of the configuration of the
water molecules around the solvent. TD-EMFT can help with the former
requirement, but the latter typically requires molecular dynamics (MD)
calculations. Previous work has shown that the effect of the solvent
on absorption spectra can be sensitive to very long-range
interactions, meaning such large system sizes are required as to make
\textit{ab initio} MD
impractical\cite{zuehlsdorff_solvent_2016}. Instead, appropriate
configurations are best obtained by conducting \textit{classical} MD
simulations of the solvated system, taking snapshots from the
resulting trajectory, carving out a section around the solute to treat
with TDDFT, and averaging the results over all
snapshots\cite{zuehlsdorff_solvent_2016,zuehlsdorff_predicting_2017}. This
procedure was followed in this work, although as the aim was not to
converge my results with respect to the number of snapshots, I
considered only a single snapshot. The structure of this snapshot is
shown in Fig.\ \ref{fig:PhPhStructures}b. The precise procedure used
to obtain this structure is detailed in Section S5 of the Supporting
Information.

I computed the absorption spectrum of phenolphthalein in water using
several different methods. Firstly, the isolated phenolphthalein
molecule in implicit solvent was treated with both the semi-local
functional PBE\cite{perdew_generalized_1996} and the hybrid functional
PBE0\cite{adamo_toward_1999}, with no embedding involved. In this
case, the PBE0 hybrid functional was used rather than B3LYP, as
previous unpublished calculations on this system using the spectral warping
approach\cite{zuehlsdorff_predicting_2017} rather than TD-EMFT (and
including significant levels of sampling) suggested that PBE0
performs better in comparison to experiment. For the PBE0 calculation,
PBE-optimised NGWFs were used, for consistency with the other
calculations.

\begin{figure}
  \centering
  \includegraphics[width=0.49\textwidth]{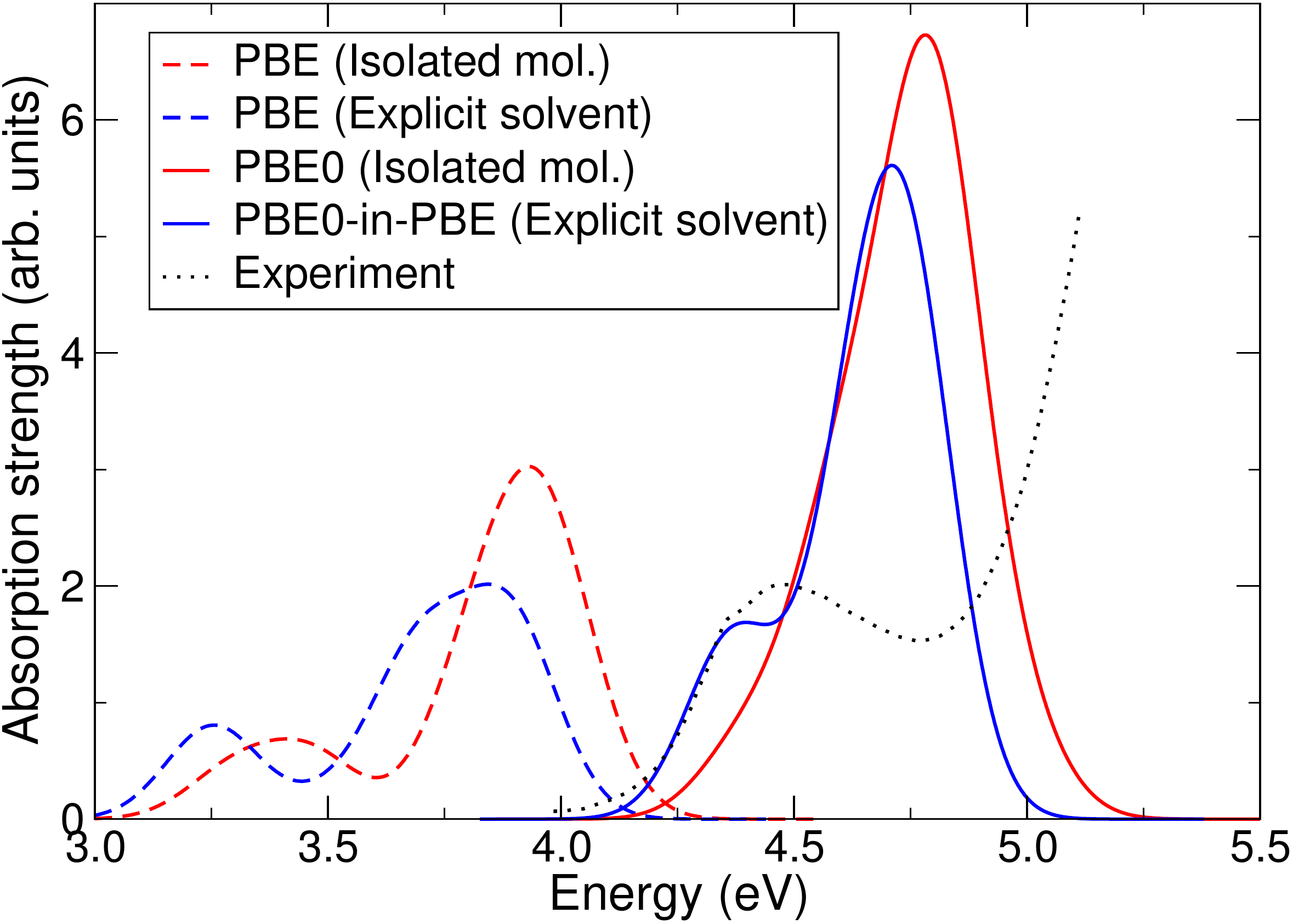}
  \caption{Absorption spectra of neutral phenolphthalein in water, as
    calculated using various different methods. The red curves
    correspond to the results obtained for the isolated molecule in
    implicit solvent only (see Fig.\ \ref{fig:PhPhStructures}a),
    whilst the blue curves correspond to the results obtained with the
    explicitly solvated system (see
    Fig.\ \ref{fig:PhPhStructures}b). Dashed lines correspond to the
    results obtained using the PBE functional, whilst solid lines
    correspond to the results obtained using the PBE0 functional (with
    PBE-optimised NGWFs) for the isolated molecule, and PBE0-in-PBE
    TD-EMFT for the explicitly solvated system, with phenolphthalein
    as the active region. The black dotted line shows the experimental
    absorption spectrum from
    Ref.\ \citenum{orndorff_absorption_1926}. All calculations are
    perfomed in implicit solvent, using the non-EMFT cavity.}
  \label{fig:PhPhImpExpSpectra}
\end{figure}

I then considered the explicitly solvated phenolphthalein system; all
calculations containing explicit solvent were also placed in implicit
solvent, giving both an explicit and an implicit layer of solvent. I
calculated the absorption spectrum of the solvated system with pure
PBE, and then with TD-EMFT, using PBE and PBE0 as the lower and higher
levels of theory respectively, with the phenolphthalein molecule as
the active region. In addition to these calculations (the results of
which are presented in Fig.\ \ref{fig:PhPhImpExpSpectra}), I performed
several others to investigate the interplay between the amount of
explicit solvent included, and the level of theory used to describe
it. Performing a full PBE0 calculation on the entire explicitly
solvated system would be extremely computationally demanding, and
therefore is not attempted here.

Norm-conserving pseudopotentials produced using the atomic solver of
the plane-wave pseudopotential DFT code
\textsc{castep}\cite{clark_first_2005} were used for all species --
details of these pseudopotentials can be found in Section S1 of the
Supporting Information. A cut-off energy of $800$~eV was used
throughout. The NGWF radii were different for the solvent molecules
and the solute itself. In the solvent molecules, H and O had valence
NGWF radii of $7$ and $9$~bohr respectively, and conduction NGWF radii
of $7$ and $11$~bohr respectively. In the solute, all NGWFs had a
radius of $11$~bohr. $4$, $4$, and $1$ NGWFs were associated with C,
O, and H atoms in all parts of the system. As noted above, all the
calculations were performed in implicit solvent, with the parameters
appropriate for water. All calculations were performed in a cubic cell
with side length $75$~bohr.

\begin{figure*}
  \centering
  \subcaptionbox{Isolated molecule}{
    \includegraphics[width=0.45\textwidth]{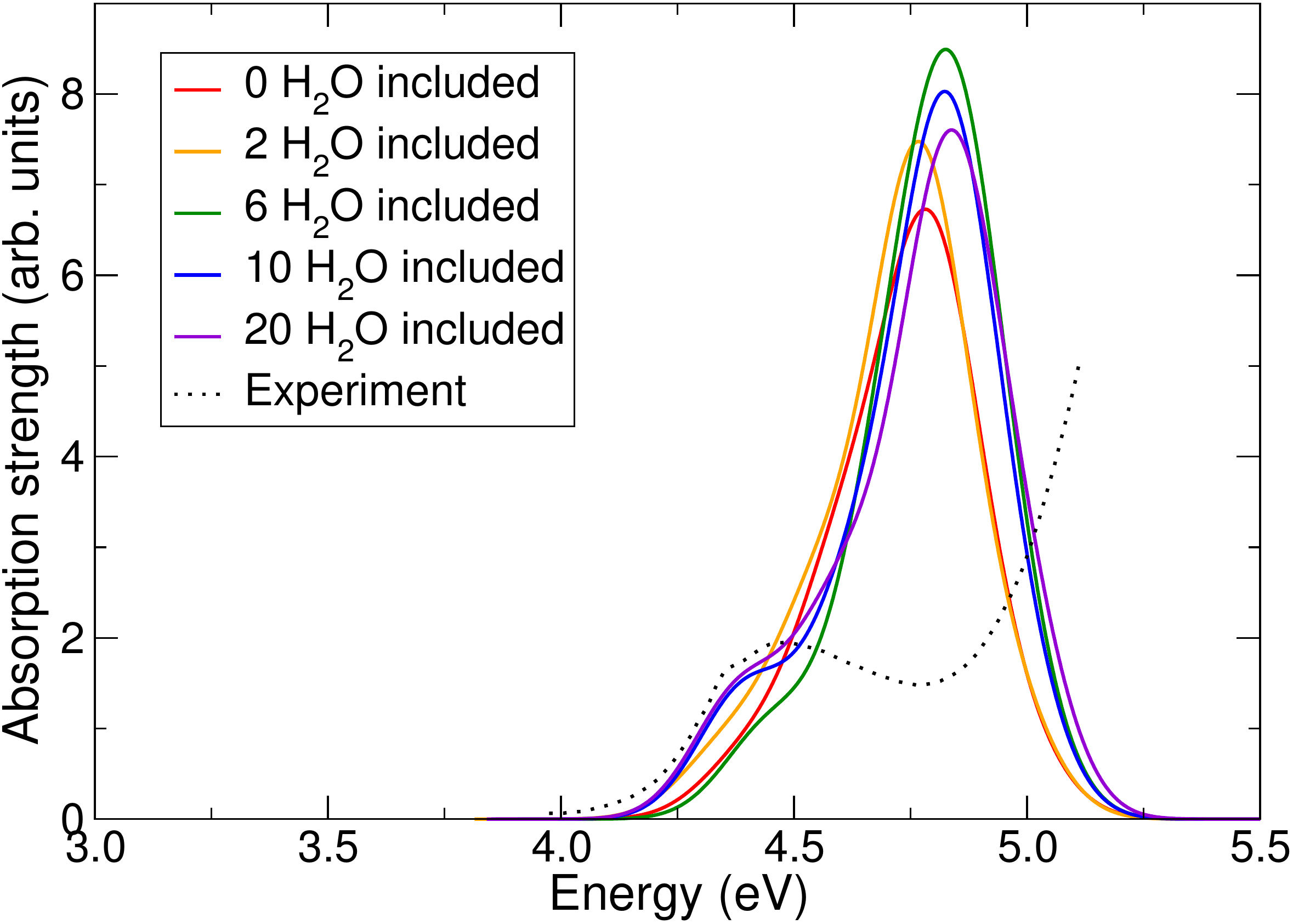}
  }
  ~
  \subcaptionbox{Solvated molecule}{
    \includegraphics[width=0.45\textwidth]{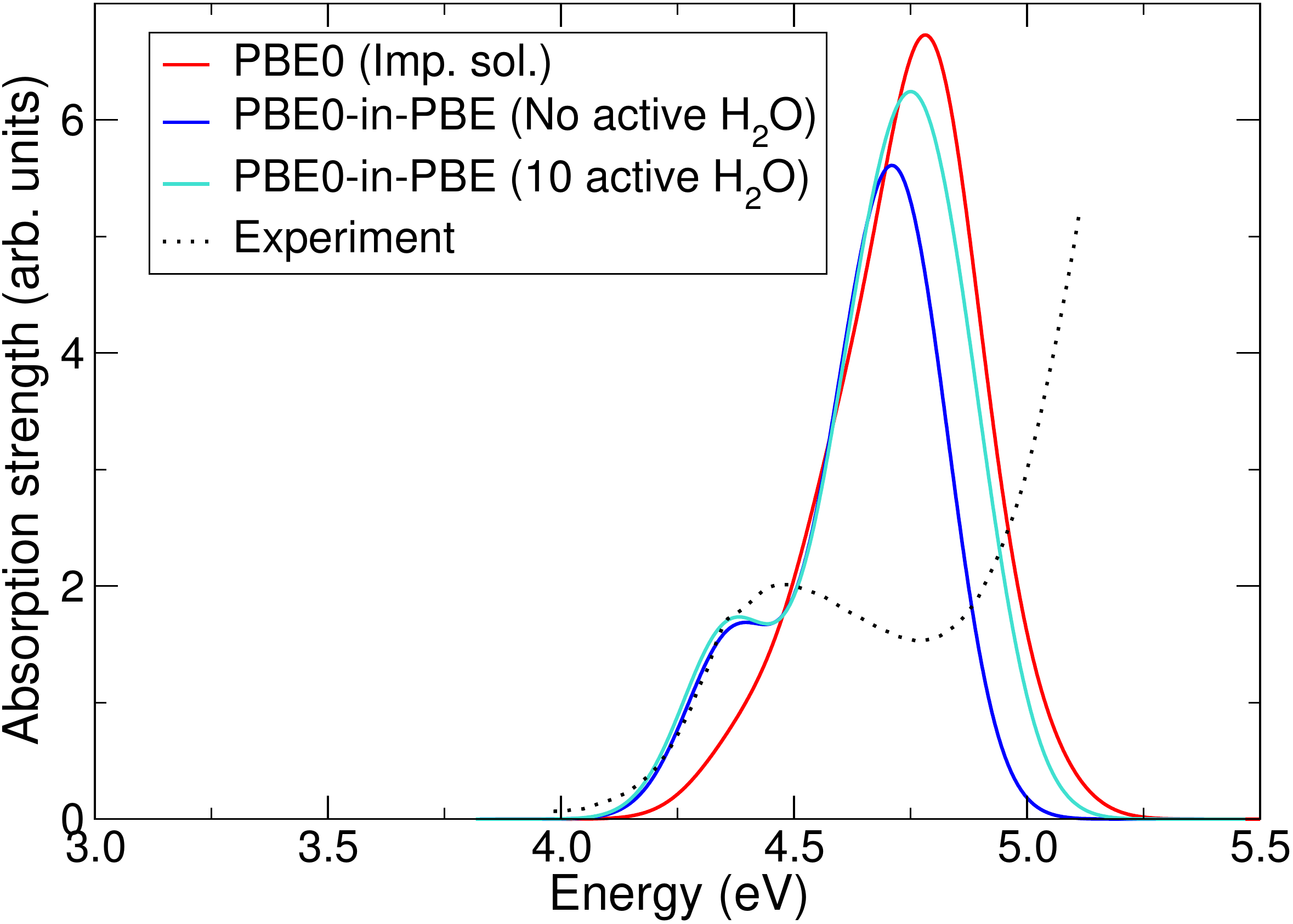}
  }
  \caption{Absorption spectra of neutral phenolphthalein in water, as
    calculated using various different methods. a) shows the
    absorption spectra calculated using the PBE0 functional (with
    PBE-optimised NGWFs) with increasing numbers of nearby water
    molecules explicitly included. The colours of the curves progress
    through the rainbow
    (red$\rightarrow$orange$\rightarrow$green$\rightarrow$blue$\rightarrow$violet)
    as the number of explicit water molecules increases. b) shows the
    absorption spectra calculated using: the PBE0 functional for
    isolated phenolphthalein; PBE0-in-PBE TD-EMFT for explicitly
    solvated phenolphthalein, with the phenolphthalein as the active
    region; and PBE0-in-PBE TD-EMFT for explicitly solvated
    phenolphthalein, with the phenolphthalein molecule and the $10$
    nearest water molecules as the active region. The first two of
    these spectra are also presented in
    Fig.\ \ref{fig:PhPhImpExpSpectra}. In both a) and b), the black
    dotted line shows the experimental absorption spectrum from
    Ref.\ \citenum{orndorff_absorption_1926}. All calculations are
    perfomed in implicit solvent, using the non-EMFT cavity.}
  \label{fig:PhPhExpWater}
\end{figure*}

Figs.\ \ref{fig:PhPhImpExpSpectra} and \ref{fig:PhPhExpWater} show the
absorption spectra calculated by the various methods detailed above,
as well as experimental data\cite{orndorff_absorption_1926} for
comparison. The experimental data exhibits two clearly separated
peaks, with the higher energy peak significantly larger than the
lower. Along with the excitation energy of these peaks, this two-peak
structure is something that calculations should replicate in order to
describe the system accurately.

The structure of the spectrum obtained with PBE (dashed lines in
Fig.\ \ref{fig:PhPhImpExpSpectra}) exhibits a two-peak structure, both
for the isolated molecule and the explicitly solvated system, although
the excitation energies of the peaks are much lower than in
experiment, as expected. This provides reassurance that the lower
level of theory (PBE) is describing the system qualitatively
correctly. The explicitly solvated system is red-shifted by around
$0.11$~eV compared to the isolated molecule, and the lower energy peak
is relatively stronger compared to the higher energy peak. Looking at
the PBE0/PBE0-in-PBE spectra (solid lines in
Fig.\ \ref{fig:PhPhImpExpSpectra}), however, it can be seen that the
two-peak structure remains for the explicitly solvated PBE0-in-PBE
calculation, but disappears for the isolated molecule PBE0 calculation
(in fact, the lower energy peak remains, but is much smaller and is
subsumed by the larger higher energy peak). The excitation energies
are now much closer to the experimental results than in the PBE case
-- in particular, the PBE0-in-PBE value for the excitation energy of
the lower peak is within $0.11$~eV of the experimental data, although
the energy of the higher peak is significantly below experiment. The
red-shift due to explicit solvation is of a similar magnitude to the
PBE case ($0.14$~eV for the higher energy peak).

The quantitative accuracy of the excitation energies could potentially
be improved by using an optimally tuned range-separated hybrid
functional rather than PBE0\cite{stein_prediction_2009}, but such
functionals are not yet available in \textsc{onetep}, so are not
considered here. It should also be noted that exact agreement with
experiment is not to be expected, as I have only looked at a single
snapshot, rather than averaging over many; however, using TD-EMFT to
calculate the absorption spectrum gives reasonable excitation
energies, whilst also maintaining the clear two-peak structure seen in
experiment. Giving a qualitatively correct description of the physics
of the system alongside reasonable quantitative predictions is not
something that is achieved by any of the other methods examined here.

To further examine the effect of including water molecules explicitly
in our calculations, and therefore demonstrating the utility of
TD-EMFT over implicit solvent or similar calculations, further
calculations treating water molecules with various levels of theory
were performed. Firstly, starting from the PBE0 isolated
phenolphthalein calculation already presented in
Fig.\ \ref{fig:PhPhImpExpSpectra}, I considered including explicit
water molecules in this calculation (also treated with PBE0), with the
molecules introduced in order of proximity to the phenolphthalein
molecule. Fig.\ \ref{fig:PhPhExpWater}a shows the absorption spectra
obtained including $0$, $2$, $6$, $10$, and $20$ water molecules in
this way. The number of water molecules was limited to a maximum of
$20$ due to the computational expense of treating more molecules with
PBE0. Secondly, I performed a PBE0-in-PBE TD-EMFT calculation on the
explicitly solvated system, similar to that already presented in
Fig.\ \ref{fig:PhPhImpExpSpectra}, but this time with the $10$ water
molecules closest to the phenolphthalein included in the active
region. The result of this calculation is presented in
Fig.\ \ref{fig:PhPhExpWater}b.

The results of Fig.\ \ref{fig:PhPhExpWater}a show that the two-peak
structure becomes more distinct as more explicit water is included --
this can be seen most extremely by comparing the isolated molecule and
explicitly solvated systems in Fig.\ \ref{fig:PhPhImpExpSpectra}, as
previously noted. The explicit inclusion of the water has a
qualitative effect on the spectrum, as seen in previous
work\cite{zuehlsdorff_solvent_2016}. The results of
Fig. \ref{fig:PhPhExpWater}b then imply that the influence of the
explicit water molecules is relatively unaffected by the level of
theory used to describe them, as there is very little difference
between the spectrum resulting from treating nearest neighbour water
molecules with PBE0, and the spectrum where only the phenolphthalein
is treated with PBE0. Taken together, Figs. \ref{fig:PhPhExpWater}a
and b form a strong argument for the utility of TD-EMFT in this system
-- including a large number of water molecules is necessary to
correctly qualitatively describe the system, but the results are
relatively insensitive to the level of (quantum mechanical) theory
used to do this, so a lower level of theory can be used. It is
important that the environment is treated quantum mechanically, rather
than classically, a point that is backed up by previous comparisons to
QM/MM methods\cite{zuehlsdorff_solvent_2016}. Overall, the results of
the calculations presented in this section demonstrate that the
implementation of TD-EMFT with implicit solvent within linear-scaling
DFT is able to successfully describe a complex system containing
several hundred atoms, giving a qualitatively correct description of
the system, and reasonably accurate quantitative results.

I also investigated the effect of using the EMFT cavity rather than
the non-EMFT cavity in the PBE0-in-PBE calculation, but found this
made effectively no difference to the results, changing the peak
excitation energies by less than $1$~meV and the oscillator strengths
by less than $0.6$\%. As outlined in Section \ref{subsec:Dimer}, this
is as expected, as the active region is now not close to the edge of
the cavity, so any change in the density kernel due to EMFT is likely
to be localised far from the cavity edge.

The results of this section also demonstrate an important point
regarding the savings TD-EMFT provides. The main limiting factor for
hybrid calculations in ONETEP is computer memory, rather than
speed. This means that the savings in memory that (TD-)EMFT provides
are as important, if not more, than any speed-up. This is demonstrated
by the fact that I was unable to reasonably perform a full hybrid
TDDFT calculation on a system containing a phenolphthalein molecule
and more than 20 explicit water molecules due to memory constraints,
but I was able to perform a calculation containing significantly more
water molecules using TD-EMFT.

\subsection{Pentacene in \textit{p}-terphenyl} \label{subsec:PentPTer}

\begin{figure*}[t]
  \centering
  \subcaptionbox{Pentacene molecule}{
    \includegraphics[trim={3cm 11cm 3cm 11cm},clip,width=0.3\textwidth]{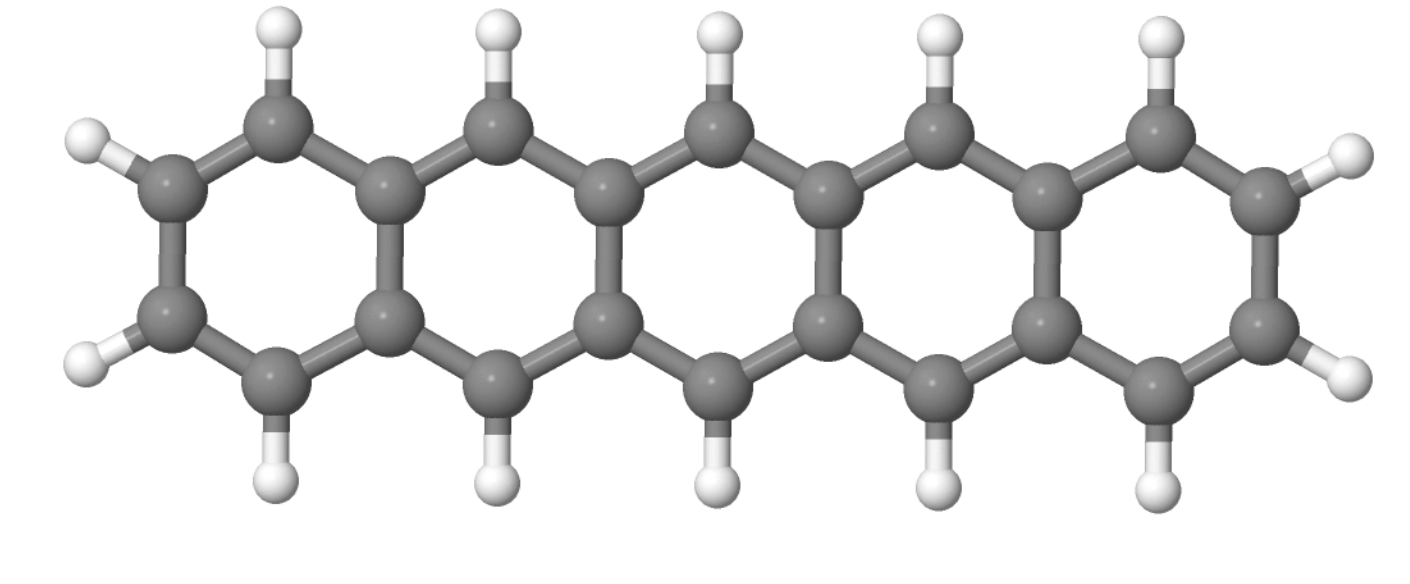}
  }
  ~
  \subcaptionbox{Cluster configuration}{
    \includegraphics[trim={4.75cm 7.75cm 4.5cm 7.75cm},clip,width=0.3\textwidth]{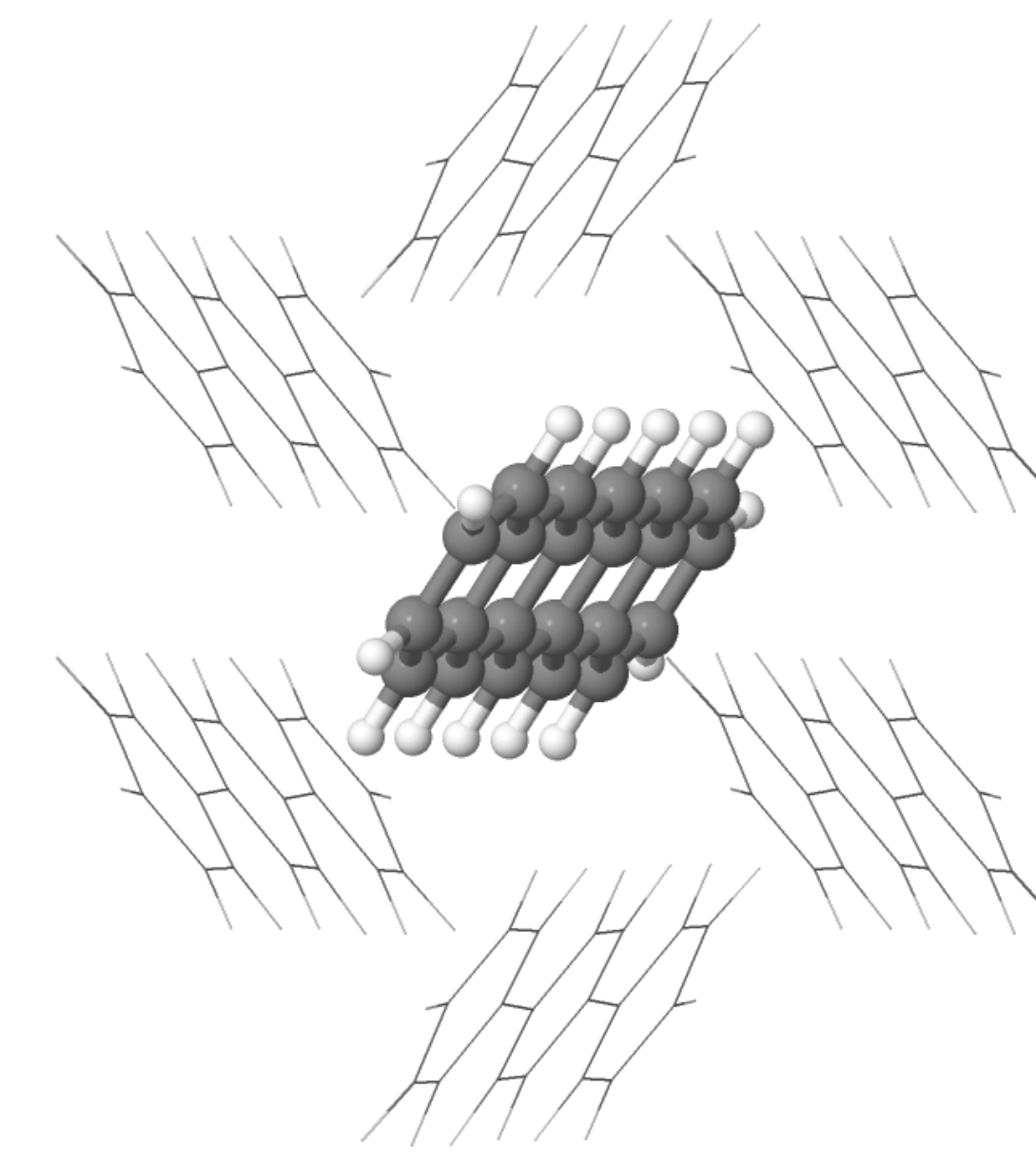}
  }
  ~
  \subcaptionbox{Crystalline configuration}{
    \includegraphics[trim={5.5cm 9.5cm 5cm 9.5cm},clip,width=0.3\textwidth]{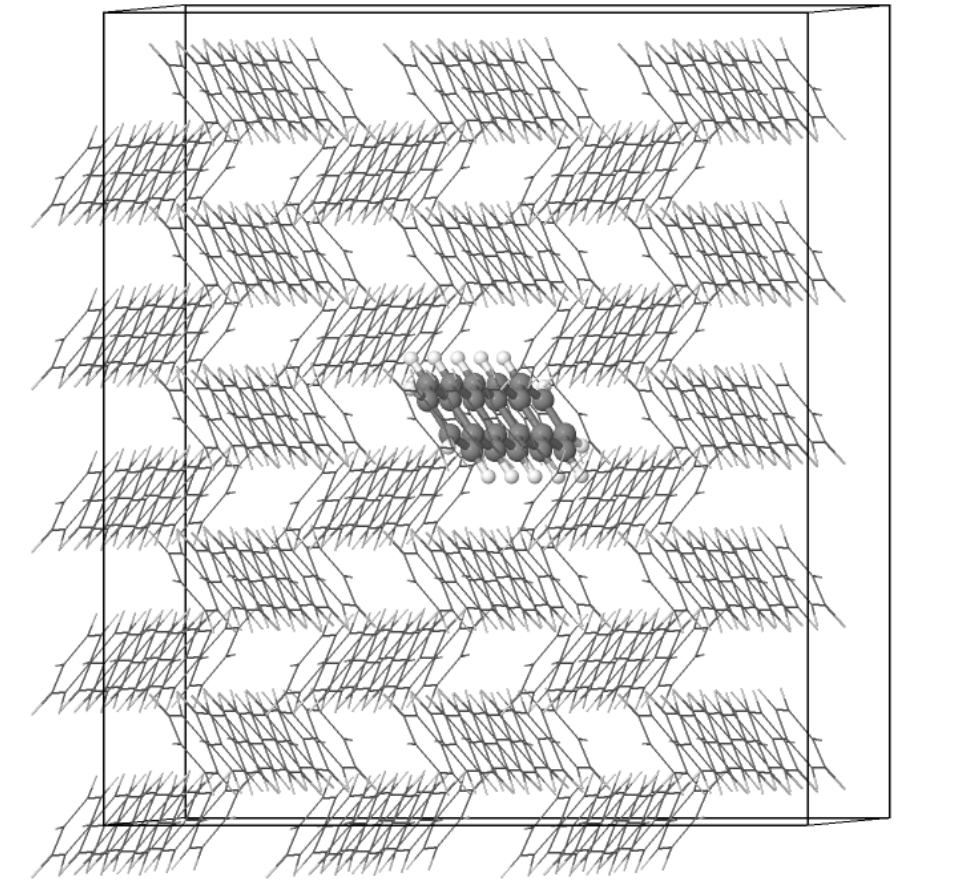}
  }
  \caption{Structures used for the calculations on the pentacene in
    \textit{p}-terphenyl system in this work. a) shows the isolated
    pentacene molecule. b) shows the cluster configuration, which is
    made up of a pentacene molecule and its six nearest neighbour
    \textit{p}-terphenyl molecules in a herring-bone structure. c)
    shows the crystalline configuration. This structure is constructed
    by taking a $3\times5\times3$ supercell of crystalline
    \textit{p}-terphenyl and replacing the central
    \textit{p}-terphenyl with a pentacene molecule. This structure is
    periodic, with the unit cell also indicated in the figure. In b)
    and c), the \textit{p}-terphenyl molecules are shown as
    wireframes, and the pentacene is shown using a ball-and-stick
    model. H and C atoms are white and grey respectively. Figures
    reproduced with permission from
    Ref.\ \citenum{prentice_combining_2020}.}
  \label{fig:PentPTerStructures}
\end{figure*}

Finally, I applied linear-scaling TD-EMFT to the pentacene-doped
\textit{para}-terphenyl molecular crystal. The author and co-workers
also studied this system in our previous work on ground state
EMFT\cite{prentice_combining_2020}, allowing comparisons to be drawn
easily. This system can be used as the basis of a room-temperature
maser\cite{oxborrow_room-temperature_2012} -- as most previously known
masing systems only work under stringent operating
conditions\cite{kleppner_properties_1962,konoplev_experimental_2006,siegman_microwave_1964},
this system has many important potential applications. Although the
population inversion necessary for masing behaviour is actually formed
between different spin states of the triplet ground state of the
pentacene molecule ($T_1$), these states are populated via a route
that starts with exciting pentacene molecules into the first excited
singlet state ($S_1$) from the (singlet) ground state
($S_0$)\cite{oxborrow_room-temperature_2012,charlton_implicit_2018}. This
means that the $S_0$ to $S_1$ transition energy ($\Delta
E_{S_0\rightarrow S_1}$) is very important, and is the focus here.

The presence of \textit{p}-terphenyl has significant effects on the
excitation energies of the pentacene
molecule\cite{kohler_intersystem_1996,patterson_intersystem_1984,zimmerman_singlet_2010,bogatko_molecular_2016,charlton_implicit_2018},
and therefore it is important to include the \textit{p}-terphenyl
environment in our calculations. The question of how much of the
environment to include is, however, a more difficult question. To this
end, three different configurations are considered, as in previous
work\cite{prentice_combining_2020} -- an isolated pentacene molecule
in vacuum, a cluster model containing the pentacene molecule and its 6
nearest \textit{p}-terphenyl neighbours, and a crystalline model
corresponding to a $3\times5\times3$ supercell of \textit{p}-terphenyl
with the central molecule substituted with pentacene. These three
structures are shown in Fig.\ \ref{fig:PentPTerStructures}a, b, and c
respectively. TD-EMFT is applied to the cluster and crystalline
structures, taking the pentacene molecule as the active region and the
surrounding \textit{p}-terphenyl molecules as the environment. The
crystalline structure in particular contains $2884$ atoms, allowing
linear-scaling TD-EMFT to be tested on a (previously unattainable)
very large system, thus including long-range interactions with the
environment.

In the solvated phenolphthalein system studied in Section
\ref{subsec:PhPh}, it was clear that a quantum mechanical description
of the environment out to long ranges, rather than a classical
description, was important. We expect this to still apply to some
degree in the crystalline system here, meaning that the full
crystalline calculation should describe the system more accurately
than a similar one using a QM/MM method. Although QM/MM methods have
been applied to the pentacene-in-\textit{p}-terphenyl system in recent
work\cite{bertoni_qm-mm_2022}, the use of semi-empirical density
functional tight binding methods for the QM region makes it difficult
to draw comparisons with the present work, as the strong dependence of
the excitations on the QM method overwhelms any effect from the
classical description of the environment.

In previous work on ground state EMFT\cite{prentice_combining_2020},
the author and co-workers were able to estimate $\Delta
E_{S_0\rightarrow S_1}$, using a combination of the $\Delta$SCF method
to obtain $\Delta E_{S_0\rightarrow T_1}$ and the Becke
method\cite{becke_singlet-triplet_2018} to obtain $\Delta
E_{T_1\rightarrow S_1}$, for each of the three structures. The $S_0$
to $S_1$ transition has also been previously studied with TDDFT: an
isolated molecule in vacuum, treated with PBE ($\Delta E_{S_0\rightarrow S_1}=1.64$~eV), B3LYP
($1.90$~eV)\cite{kadantsev_electronic_2006}, and an optimally tuned
range-separated hybrid functional (OT-LC$\omega$PBE,
$2.15$~eV)\cite{charlton_implicit_2018}; an isolated molecule in
\textit{p}-terphenyl-like implicit solvent, treated with PBE
($1.60$~eV) and OT-LC$\omega$PBE
($2.07$~eV)\cite{charlton_implicit_2018}, or treated with PBE and
empirically corrected ($2.27$~eV)\cite{bogatko_molecular_2016}; and
the cluster configuration, treated with PBE ($1.58$~eV) and
OT-LC$\omega$PBE ($2.09$~eV)\cite{charlton_implicit_2018}. The results
of this work can be compared to these previous results, and also to
experimental data\cite{heinecke_laser_1998,kohler_intersystem_1996}.

Here, PBE is used as the lower level of theory, with B3LYP as the
higher level. The norm-conserving pseudopotentials distributed with
\textsc{onetep} were used for both species, the cut-off energy was
taken as $750$~eV, and the NGWF radii were set to $11$~bohr for all
atoms, with 4 NGWFs associated with the C atoms, and 1 with the H
atoms. All three configurations were performed in PBCs, with the same
unit cell (that of the $3\times5\times3$ supercell of
\textit{p}-terphenyl).

\begin{table*}[t]
  \centering
  \begin{tabular}{ | c | c c c | }
      \cline{2-4}
      \multicolumn{1}{ c }{} & \multicolumn{3}{ | c | }{$\Delta E_{S_0\rightarrow S_1}$ (eV)} \\
      \hline
      Configuration & PBE & B3LYP-in-PBE & Exp. \\
      \hline
      Vacuum & $1.880$ & $2.198$ & $2.31$\cite{heinecke_laser_1998} \\
      Cluster & $1.792$ & $2.069$ & \multirow{2}{*}{$2.09$\cite{kohler_intersystem_1996}} \\
      Crystal & $1.810$ & $2.089$ & \\
      \hline
  \end{tabular}
  \caption{Excitation energies calculated using TDDFT/TD-EMFT for the
    transition between the $S_0$ and $S_1$ states for pentacene, in
    the three geometries shown in
    Fig.\ \ref{fig:PentPTerStructures}. Experimental data from
    Refs.\ \citenum{heinecke_laser_1998,kohler_intersystem_1996} are
    also shown. For the cluster and crystal configurations,
    B3LYP-in-PBE refers to a TD-EMFT calculation, whereas for the
    vacuum configuration, it corresponds to a B3LYP calculation
    performed using PBE-optimised NGWFs.} \label{tab:PentPTerSpectra}
\end{table*}

Table \ref{tab:PentPTerSpectra} presents the results of the
calculations on the three structures, alongside experimental
data\cite{heinecke_laser_1998,kohler_intersystem_1996}. Fig.\ S3 in
Section S6 of the Supporting Information shows the absorption spectra
corresponding to the same data. It can immediately be seen that the
B3LYP-in-PBE TD-EMFT calculations for both the cluster and crystal
configurations match well with crystalline experimental data -- in
fact, the crystalline calculation match experiment almost exactly,
with the cluster calculation $0.02$~eV lower. This demonstrates the
importance of including long-range interactions between the pentacene
and its environment. The ordering of the three configurations in terms
of excitation energy (cluster, crystal, vacuum) and in terms of
absorption strength (crystal, vacuum, cluster) is the same for both
PBE and B3LYP/B3LYP-in-PBE. These results provide evidence that
linear-scaling TD-EMFT is correctly describing the excitation, and
gives quantitatively accurate results for systems containing thousands
of atoms. The B3LYP-in-PBE TD-EMFT calculations also produce values
for $\Delta E_{S_0\rightarrow S_1}$ significantly closer to experiment
than the indirect method used in previous work ($1.85$ and $1.76$~eV
for the cluster and crystal configurations
respectively)\cite{prentice_combining_2020}. This demonstrates the
utility of using TD-EMFT directly, rather than only ground state EMFT.

It should also be noted that the vacuum calculations, including those
done with B3LYP, underestimate the value measured experimentally in
vacuum. This is in line with previous computations performed with
hybrid DFT, even when using an optimally tuned range-separated hybrid
functional\cite{charlton_implicit_2018}. Higher-order methods such as
multi-reference M{\o}ller-Plesset perturbation theory do give the
correct excitation energy\cite{zeng_low-lying_2014}, but are not
implemented within the EMFT framework within \textsc{onetep}, and are
therefore not considered here. Hybrid functionals such as B3LYP
provide a more reliable description of the excitation spectrum in the
solid state, where screening reduces the HOMO-LUMO gap, bringing it in
line with hybrid functional
predictions\cite{refaely-abramson_gap_2013}.

\section{Conclusions} \label{sec:Conclusions}

In this work, I have presented the first implementation of
time-dependent embedded mean field theory combined with both
linear-scaling density functional theory and a classical implicit
solvation model, all within the linear-scaling DFT code
\textsc{onetep}. This combination allows for multi-level simulations
of electronic excitations of large-scale systems to be conducted, with
two levels of DFT and a classical continuum model all contained within
the same calculation. Such calculations will likely be extremely
useful in systems where excitations of interest are largely localised
on a particular active region, but the environment affects these
excitations both quantum mechanically and classically. I have demonstrated
the power and utility of this method by applying it to a wide range of
different systems, including the water-nitrogen molecular dimer,
phenolphthalein in water, and pentacene-doped $p$-terphenyl. In each
case, the linear-scaling TD-EMFT method obtains excellent results, agreeing
well with experimental data and previous calculations. These
calculations also demonstrated that the method can be used for systems
containing thousands of atoms, which would not have previously been
accessible for purely high-accuracy hybrid functional TDDFT. This work
will allow embedding calculations of electronic excitations to be
applied to an even wider range of problems than previously, both in
terms of scale and also in terms of systems of interest in physics,
chemistry, and materials science.

\begin{suppinfo}

  Further details of pseudopotentials used in this work; outline of
  derivation of TDDFT equations and algorithm to solve them as used in
  this work; additional absorption spectra data for water-nitrogen
  dimer; further details on method for obtaining snapshot used for
  phenolphthalein in water calculations; absorption spectra for
  pentacene in \textit{p}-terphenyl; \texttt{.cif} files for
  structures used.

\end{suppinfo}

\begin{acknowledgement}
  The author acknowledges the support of St Edmund Hall, University of
  Oxford, through the Cooksey Early Career Teaching and Research
  Fellowship. The author is grateful to the UK Materials and Molecular
  Modelling Hub for computational resources, which is partially funded
  by EPSRC (EP/P020194/1 and EP/T022213). Computational resources were
  also provided by the ARCHER/ARCHER2 UK National Supercomputing
  Service, for which access was obtained via the UKCP consortium
  (EP/P022561/1). The author is also grateful to Prof.\ Arash Mostofi
  and Laura Prentice for useful comments on the manuscript.
\end{acknowledgement}


\bibliography{EMFTBib}

\end{document}


\title{Supporting Information for \\ Efficiently computing excitations of complex systems: linear-scaling time-dependent embedded mean-field theory in implicit solvent}

\author{Joseph C.\ A.\ Prentice}

\maketitle

\renewcommand\thefigure{S\arabic{figure}}
\renewcommand\thesection{S\arabic{section}}
\renewcommand{\bibnumfmt}[1]{S#1}
\renewcommand{\citenumfont}[1]{S#1}

\section{Psuedopotentials} \label{sec:Pseudopotentials}

The pseudopotential strings defining the pseudopotentials used in the main text for the calculations on phenolphthalein are:
\begin{itemize}
\item C: 1|1.2|17|20|23|20N:21L(qc=8)
\item O: 1|1.2|23|26|31|20N:21L(qc=9)
\item H: 1|0.8|14|16|19|10N(qc=8)
\end{itemize}

\section{Derivation of TDDFT equations} \label{sec:TDDFTDerivation}

If the ground state of the system is perturbed by a small
time-dependent perturbation with frequency $\omega$, $\delta v
(\mathbf{r},\omega)$, the change in the density $\rho(\mathbf{r})$ to
first order (the linear response) will be
\begin{equation}
  \delta\rho(\mathbf{r},\omega) = \int d^3\mathbf{r}' \, \chi(\mathbf{r},\mathbf{r}',\omega) \delta v(\mathbf{r}',\omega) ~.
  \label{eq:ResponseDens}
\end{equation}
Here, $\chi$ is the density-density linear-response function
\begin{equation}
  \chi(\mathbf{r},\mathbf{r}',\omega) = \sum_{v,c} \left[ \frac{\psi_v^*(\mathbf{r})\psi_c(\mathbf{r})\psi_v(\mathbf{r}')\psi_c^*(\mathbf{r}')}{\omega - (\epsilon_c - \epsilon_v)} - \frac{\psi_v(\mathbf{r})\psi_c^*(\mathbf{r})\psi_v^*(\mathbf{r}')\psi_c(\mathbf{r}')}{\omega + (\epsilon_c - \epsilon_v)} \right] ~, \label{eq:LinearResponseFn}
\end{equation}
where the $\psi_v$ and $\psi_c$ are the valence and conduction
ground-state Kohn-Sham eigenstates respectively, with $\epsilon_v$
and $\epsilon_c$ as their corresponding eigenvalues. $\delta v$ is
defined self-consistently as\cite{ding_linear-response_2017}
\begin{equation}
  \delta v(\mathbf{r},\omega) = \delta v_{\text{ext}} (\mathbf{r},\omega) + \delta v_H (\mathbf{r},\omega) + \int d^3\mathbf{r}' \, f_{xc} (\mathbf{r},\mathbf{r}',\omega) \delta\rho(\mathbf{r}',\omega) ~, \label{eq:TDDFTDeltav}
\end{equation}
where $\delta v_{\text{ext}}$ and $\delta v_H$ represent the change in
the external and Hartree potentials respectively. $f_{xc}$ is often
known as the exchange-correlation
kernel\cite{zuehlsdorff_linear-scaling_2013}, and is given by the
second derivative of the exchange-correlation energy with respect to
density. Taking the adiabatic
approximation\cite{ding_linear-response_2017,maitra_perspective_2016},
means that the frequency dependence of $f_{xc}$ can be neglected,
resulting in
\begin{equation}
  f_{xc}(\mathbf{r},\mathbf{r}') = \frac{\delta^2 E_{xc} [\rho]}{\delta\rho(\mathbf{r}) \delta\rho(\mathbf{r}')} ~. \label{eq:XCKernel}
\end{equation}
With these definitions, it is a simple task to follow the derivation
set out in previous
work\cite{maitra_perspective_2016,casida_progress_2012,zuehlsdorff_linear-scaling_2015},
leading to the Casida formulation of the TDDFT eigenvalue equation:
\begin{equation}
  \begin{pmatrix} A & B \\
    -B & -A \end{pmatrix} \begin{pmatrix} X \\ Y \end{pmatrix} = \omega \begin{pmatrix} X \\ Y \end{pmatrix} ~. \label{eq:CasidaTDDFT}
\end{equation}
Here, $A$ and $B$ are matrices, that can be written in a basis of the
valence and conduction ground-state Kohn-Sham eigenstates as
\begin{equation}
  A_{cv,c'v'} = \delta_{c,c'} \delta_{v,v'} (\epsilon_c - \epsilon_v) + Q_{cv,c'v'} ~,
\end{equation}
and
\begin{equation}
  B_{cv,c'v'} = Q_{cv,c'v'} = \iint d^3\mathbf{r} \, d^3\mathbf{r}' \, \psi^*_c(\mathbf{r}) \psi^*_v(\mathbf{r}) \left[ \frac{1}{|\mathbf{r}-\mathbf{r}'|} + f_{xc}(\mathbf{r},\mathbf{r}') \right] \psi_{c'}(\mathbf{r}') \psi_{v'}(\mathbf{r}') ~. \label{eq:CouplingMatDef}
\end{equation}

The two parts of the eigenvector part of Eq.\ \eqref{eq:CasidaTDDFT},
$X$ and $Y$, represent excitation and de-excitation processes
respectively\cite{zuehlsdorff_linear-scaling_2015}. The coupling
between these two processes can be neglected by assuming that the
off-diagonal blocks $B$ are zero (although the contribution of the
coupling matrix $Q$ to the on-diagonal blocks $A$ is still
included). This is known as the Tamm-Dancoff approximation
(TDA)\cite{hirata_time-dependent_1999}, and has the advantage of
reducing the non-Hermitian eigenvalue problem in
Eq.\ \eqref{eq:CasidaTDDFT} to a Hermitian one:
\begin{equation}
  AX = \omega X \label{eq:TDATDDFT} ~.
\end{equation}
As noted in the main text, the TDA usually gives reliable excitation
frequencies, but can give significant errors in some situations, and
performs more poorly for the computation of oscillator
strengths\cite{zuehlsdorff_linear-scaling_2015}. Because of the
reduction in complexity, and thus computational cost, the TDA is used
throughout this work.

\section{Algorithm for TDDFT calculations in ONETEP} \label{sec:TDDFTAlgorithm}

To calculate the $N_\omega$ lowest excitation frequencies of the
system, the function
\begin{equation}
  \Omega = \sum_i^{N_\omega} \omega_i = \sum_i^{N_\omega} \frac{\textrm{Tr}\left(P_i^{\{1\}\dagger}S^c q_i S^v\right)}{\textrm{Tr}\left(P_i^{\{1\}\dagger}S^c P_i^{\{1\}} S^v\right)}
\end{equation}
must be minimised with respect to the response density matrices
$P_i^{\{1\}}$, whilst constraining the TDDFT eigenvectors to be
orthonormal
\begin{equation}
  \textrm{Tr}\left(P_i^{\{1\}\dagger}S^c P_j^{\{1\}} S^v\right) = \delta_{ij} ~.
\end{equation}
Here, $S^c$ and $S^v$ are the overlap matrices for the set of
conduction NGWFs and valence NGWFs respectively. The response density
matrices are the TDDFT eigenvectors ($X$ in Eq.\ \eqref{eq:TDATDDFT})
expressed in terms of the NGWFs:
\begin{equation}
  P_i^{\{1\}\alpha\beta} = \sum_{cv} \braket{\phi^\alpha|\psi_c} X^i_{cv} \braket{\psi_v|\phi^\beta} ~.
\end{equation}
$q_i$ represents the action of the TDDFT operator on a trial response
density matrix, given by
\begin{equation}
  q_i^{\alpha\beta} = \left( K_c H^c P_i^{\{1\}} - P_i^{\{1\}} H^v K_v \right)^{\alpha\beta} + \left(K_c V_{\text{SCF}i}^{\{1\}} K_v \right)^{\alpha\beta} ~,
\end{equation}
where $H^v$ is the ground-state Hamiltonian, $H^c$ is the projected
Hamiltonian used in the conduction NGWF optimisation, and $K_v$ and
$K_c$ are the corresponding valence and conduction density
kernels. $V_{\text{SCF}i}^{\{1\}}$ is the potential generated by the
response density, given by
\begin{equation}
  \left(V_{\text{SCF}i}^{\{1\}}\right)_{\alpha\beta} = 2 \iint d^3\mathbf{r} d^3\mathbf{r}' \, \phi^*_\alpha(\mathbf{r}) \phi_\beta(\mathbf{r}) \rho^{\{1\}}_i(\mathbf{r}') \left[ \frac{1}{|\mathbf{r}-\mathbf{r}'|} + f_{xc}(\mathbf{r},\mathbf{r}') \right] ~, \label{eq:V_SCF}
\end{equation}
where $\rho^{\{1\}}_i(\mathbf{r}) = \sum_{cv} \psi_c(\mathbf{r})
X^i_{cv} \psi_v(\mathbf{r})$. If a hybrid functional is being used, a
contribution from exact exchange should be added to
$V_{\text{SCF}i}^{\{1\}}$, given
by\cite{zuehlsdorff_linear-scaling_2013,dziedzic_linear-scaling_2013}
\begin{equation}
  \left(V_{\text{SCF}i}^{\{1\}\text{XX}}\right)^{\alpha\beta} = -2 f_{\text{XX}} \sum_{\gamma\delta} P_i^{\{1\}\gamma\delta} \left( \alpha \gamma | \delta \beta \right) ~. \label{eq:V_SCFXX}
\end{equation}
In order to conduct a TD-EMFT calculation, this procedure is modified
simply by replacing $f_{xc}$ in Eq.\ \eqref{eq:V_SCF} with the
expression given in Eq.\ (6) in the main text. If
a hybrid functional is being used in the active region, an exact
exchange contribution should also be added, but with the exchange
restricted to be within the active region only, modifying
Eq.\ \eqref{eq:V_SCFXX} to become
\begin{equation}
  \left(V_{\text{SCF}i}^{\{1\}\text{XX}}\right)^{\alpha\beta} = -2 f_{\text{XX}} \sum_{\gamma\delta\in A} P_i^{\{1\}\gamma\delta} \left( \alpha \gamma | \delta \beta \right) ~.
\end{equation}

\section{Higher energy absorption spectra for water-nitrogen dimer} \label{sec:WaterN2MoreEx}

\begin{figure}
  \centering
  \includegraphics[width=0.5\textwidth]{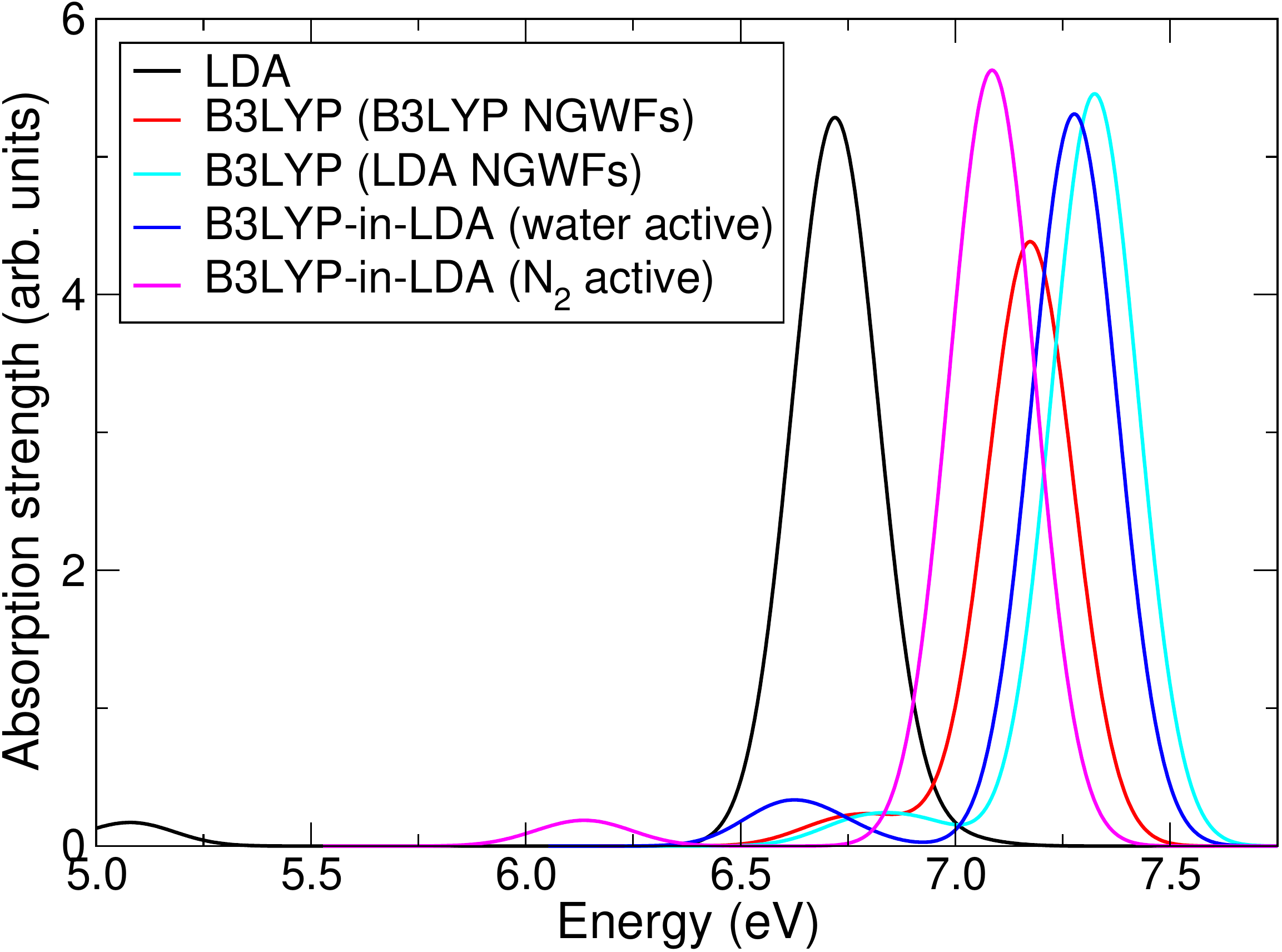}
  \caption{Absorption spectra of the water-nitrogen dimer, including
    higher energy excitations than those seen in Figs.\ 2 and 3 in the
    main text, calculated at various levels of theory: full system LDA
    (black), full system B3LYP with LDA-optimised NGWFs (cyan), full
    system B3LYP with B3LYP-optimised NGWFs (red), B3LYP-in-LDA with
    water as the active region (blue), and B3LYP-in-LDA with nitrogen
    as the active region (magenta). These results were calculated in
    vacuum.}
  \label{fig:WaterN2DimerMoreExSpectra}
\end{figure}

\begin{figure*}
  \centering
  \subcaptionbox{LDA}{
    \includegraphics[trim={15cm 3cm 14cm 3cm},clip,width=0.45\textwidth]{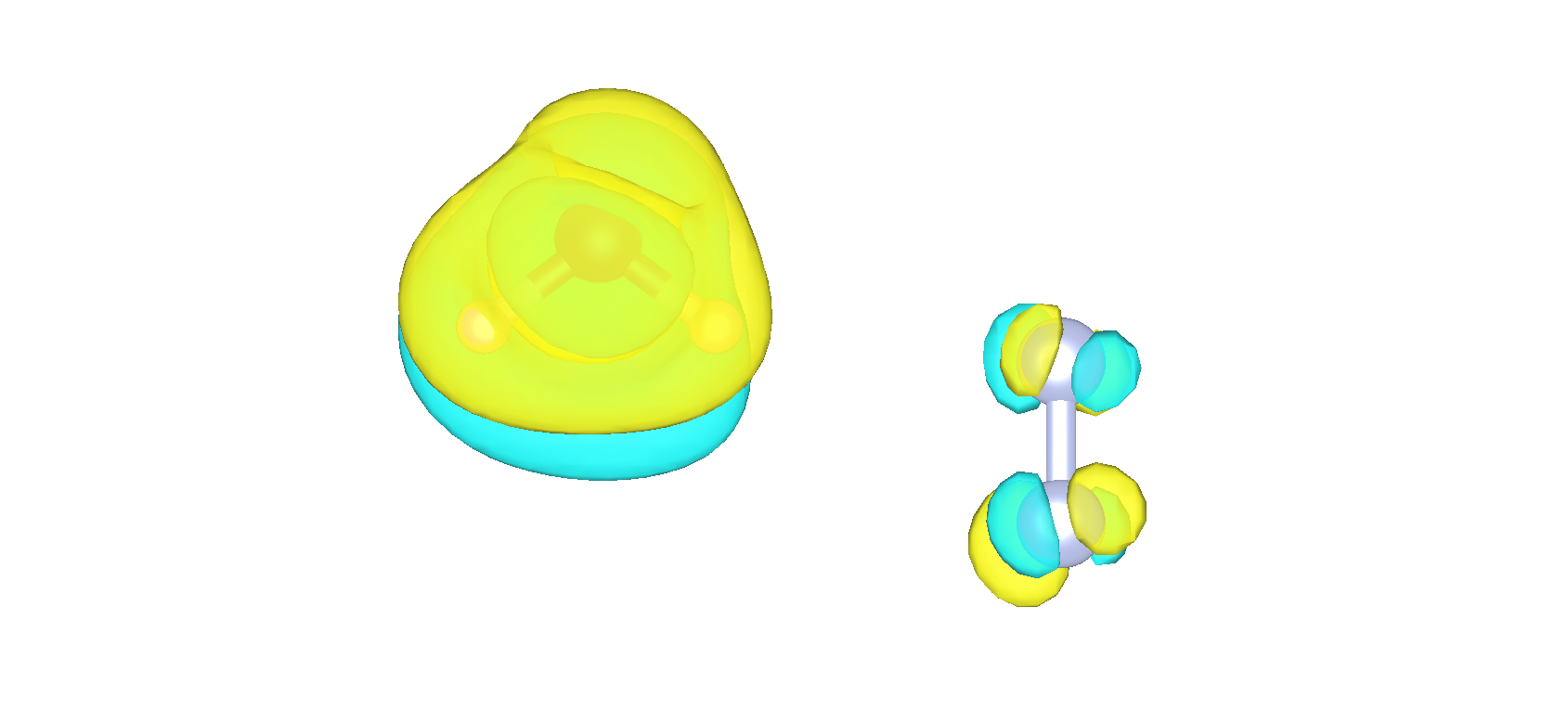}
  }
  ~
  \subcaptionbox{B3LYP}{
    \includegraphics[trim={15cm 3cm 14cm 3cm},clip,width=0.45\textwidth]{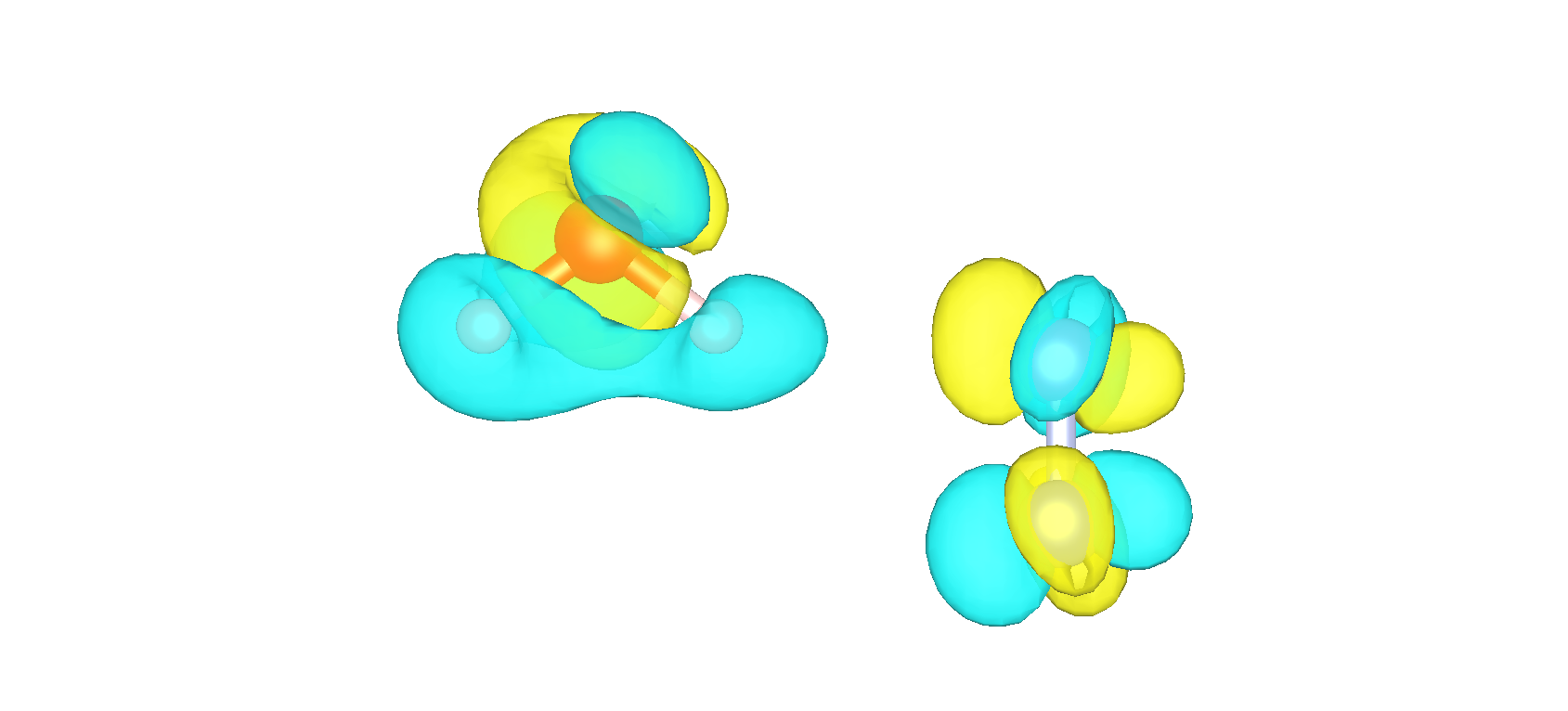}
  }
  ~
  \subcaptionbox{B3LYP-in-LDA (water active)}{
    \includegraphics[trim={15cm 3cm 14cm 3cm},clip,width=0.45\textwidth]{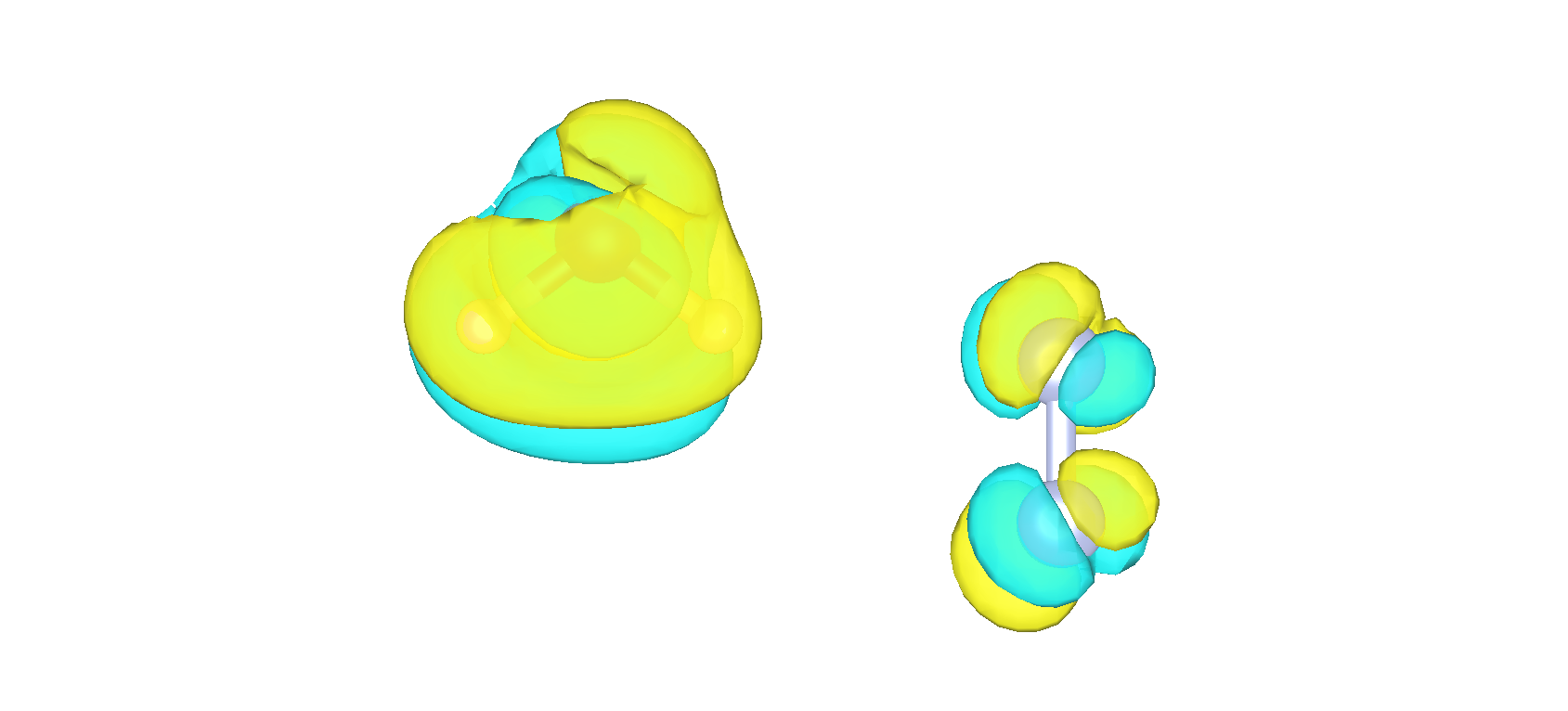}
  }
  ~
  \subcaptionbox{B3LYP-in-LDA (N$_2$ active)}{
    \includegraphics[trim={15cm 3cm 14cm 3cm},clip,width=0.45\textwidth]{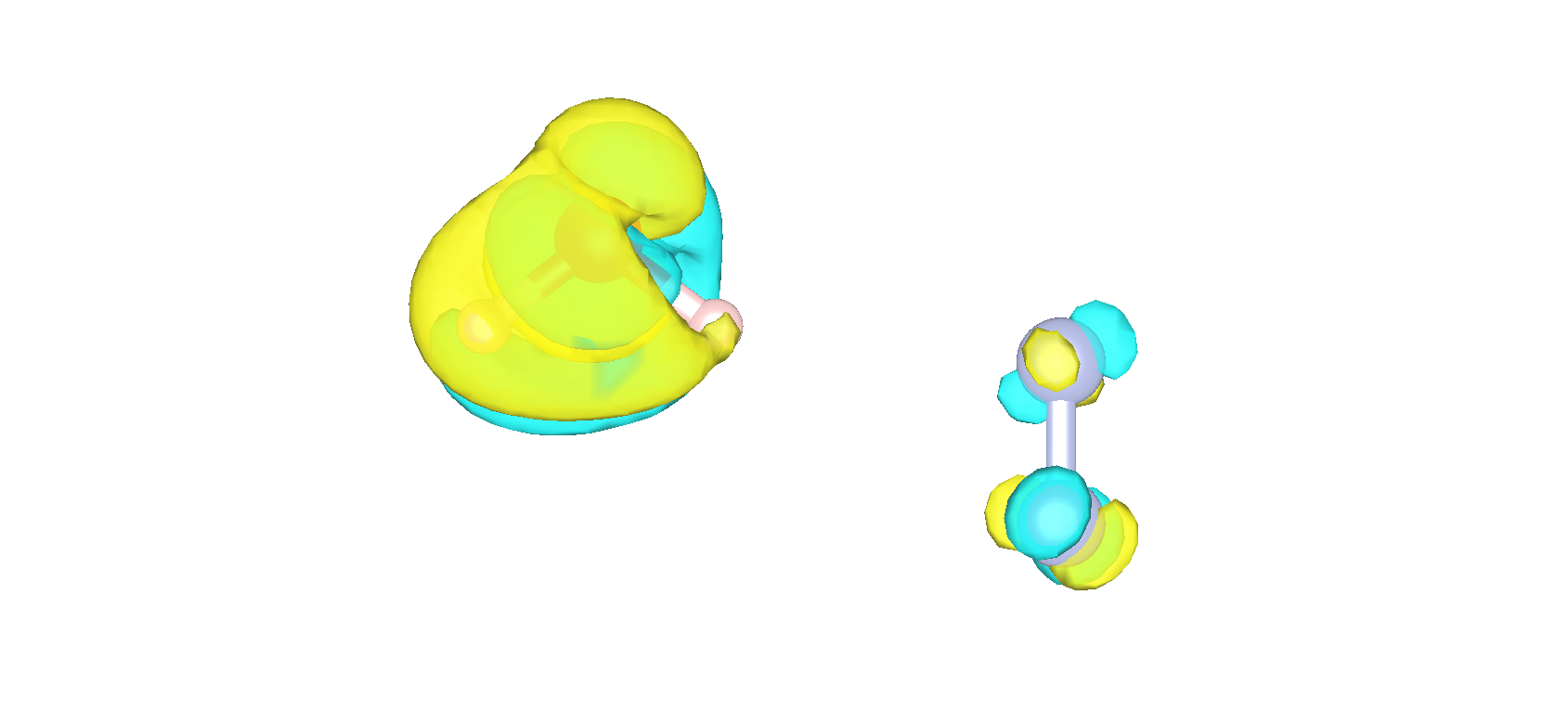}
  }
  \caption{Isosurfaces of the calculated response density for the
    strongest excitation seen in
    Fig.\ \ref{fig:WaterN2DimerMoreExSpectra}, calculated at various
    levels of theory: full system LDA, full system B3LYP (with
    B3LYP-optimised NGWFs), B3LYP-in-LDA with water as the active
    region, and B3LYP-in-LDA with nitrogen as the active region. All
    response densities were computed in vacuum. The isosurfaces are at
    $|n|=0.01$~e\,\r{A}$^{-3}$, with yellow and blue representing
    positive and negative response densities respectively. O, N, and H
    atoms are red, blue, and white respectively. Figures produced
    using VESTA\cite{momma_vesta_2011}.}
  \label{fig:WaterN2DimerMoreExCharacter}
\end{figure*}

Fig.\ \ref{fig:WaterN2DimerMoreExSpectra} presents the absorption
spectra of the water-nitrogen dimer including the higher energy bright
state discussed in the main text, as calculated in vacuum using a
variety of different methods -- this is directly comparable to
Fig.\ 2a in the main text. Fig.\ \ref{fig:WaterN2DimerMoreExCharacter}
presents isosurfaces of the response density corresponding to this
excitation -- this is comparable to Fig.\ 4 in the main text, although
the excitations shown here were computed in vacuum, rather than in
implicit solvent. As noted in the main text, LDA does a significantly
better job of describing this higher energy excitation relative to
B3LYP than excitations of lower energy -- the difference between the
LDA and full B3LYP excitation energies is $0.45$~eV.

The key difference to notice in
Fig.\ \ref{fig:WaterN2DimerMoreExCharacter} compared to Fig.\ 4 in the
main text is that the nitrogen molecule is now more involved in the
excitation, particularly in the B3LYP results, although the water
molecule is still more dominant. B3LYP and LDA calculations result in
somewhat different characters, with the characters of the B3LYP-in-LDA
calculations somewhere in between. This enables us to make sense of
the results in Fig.\ \ref{fig:WaterN2DimerMoreExSpectra}. The
B3LYP-in-LDA result with water as the active region is closer to the
full B3LYP result than the B3LYP calculation performed with
LDA-optimised NGWFs (the errors are $0.10$ and $0.15$~eV
respectively). This can be understood as the description of the water
molecule with B3LYP giving most of the `correct' result, with the LDA
description of the nitrogen molecule providing a contribution that
pulls the excitation downwards in energy. The reverse is true for the
B3LYP-in-LDA calculation with nitrogen as the active region --
compared to the lower energy excitation discussed in the main text,
the error for this calculation is somewhat smaller, as the nitrogen
molecule is more involved in the excitation, and LDA gives a better
description compared to B3LYP.

\section{Obtaining explicitly solvated structure for phenolphthalein in water} \label{sec:AMBER}

To obtain the structure shown in Fig.\ 5b, I
used the classical MD code \textsc{amber}
(v.\ 16)\cite{salomon-ferrer_overview_2013}. First, I used
\textsc{amber} to generate a classical force field for the
phenolphthalein molecule. This force field was of the GAFF
form\cite{wang_development_2004}, and the parameters for it were
generated using \textsc{amber}'s own \texttt{antechamber} tool, using
the AM1-BCC method\cite{jakalian_fast_2000} to assign charges. The
force field's description of the most important degrees of freedom of
the system then needed to be validated against DFT calculations. In
this case, I investigated the dihedral rotation of the phenol group,
and found that the potential energy surface predicted by the classical
force field matched that predicted by DFT up to energies corresponding
to $300$~K, which was adequate for the purposes of this work.

After obtaining a force field for the phenolphthalein molecule, I
solvated it in around $11000$ water molecules, described using the
TIP3P model\cite{mark_structure_2001}. The energy of this system was
then minimised, before being heated from $0$ to $300$~K over
$20$~ps. The volume of the system was then allowed to equilibrate in
the NPT ensemble at $1$~atm and $300$~K for $400$~ps, before finally
performing a production run of $8$~ns in the NVT ensemble. The
Langevin thermostat was used through, with a collision frequency of
$1$~ps$^{-1}$, and a time step of $2$~fs was also used throughout. To
enable the use of such a relatively large time step, I constrained all
bonds involving hydrogen using the SHAKE
algorithm\cite{ryckaert_numerical_1977,elber_shake_2011}. The snapshot
was then extracted from the trajectory obtained in the production run.

\section{Pentacene in \textit{p}-terphenyl absorption spectra} \label{sec:PentPTerSpectra}

\begin{figure}
  \centering
  \includegraphics[width=0.49\textwidth]{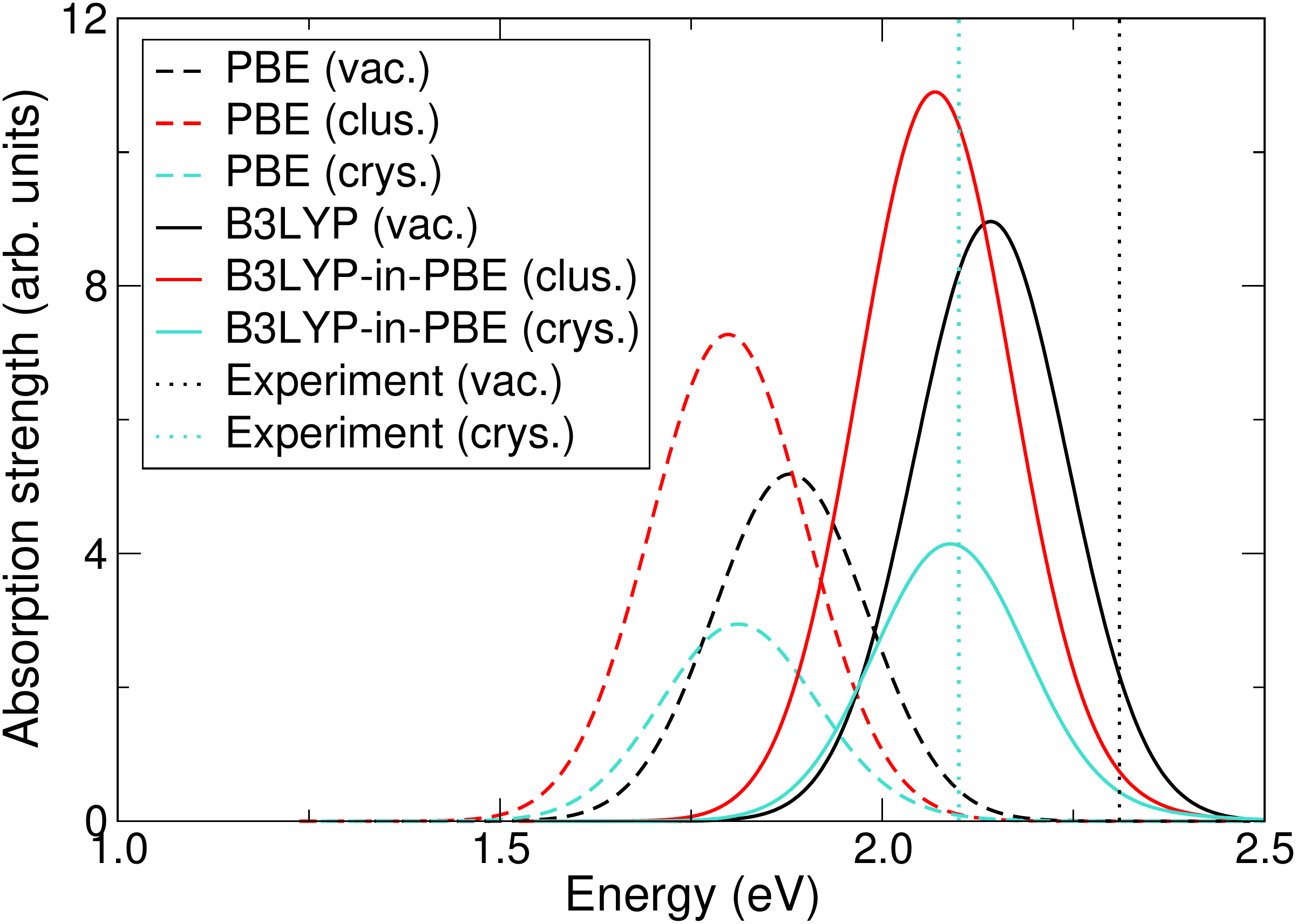}
  \caption{Absorption spectra of pentacene in the three different
    configurations shown in Fig.\ 8, as calculated using PBE alone,
    B3LYP alone, or B3LYP-in-PBE TD-EMFT. Previous experimental data
    are also shown for reference. Black lines show the spectra
    calculated for isolated pentacene in vacuum (Fig.\ 8a in the main
    text). Red lines show the spectra calculated for pentacene within
    a cluster of $6$ \textit{p}-terphenyl molecules (Fig.\ 8b in the
    main text). Turquoise lines show the spectra calculated for
    pentacene embedded within a \textit{p}-terphenyl crystal, with a
    periodic cell containing $89$ $p$-terphenyl molecules (Fig.\ 8c in
    the main text). Dashed lines show the spectra calculated using PBE
    only. Solid lines show the spectra calculated with B3LYP only (for
    isolated pentacene in vacuum) or B3LYP-in-PBE TD-EMFT (for the
    cluster and crystal configurations). The black dotted vertical
    line marks the experimentally measured energy of the absorption
    peak for pentacene in vacuum, taken from
    Ref.\ \citenum{heinecke_laser_1998}; the turquoise dotted vertical
    line marks the same quantity for pentacene embedded within
    \textit{p}-terphenyl, taken from
    Ref.\ \citenum{kohler_intersystem_1996}.}
  \label{fig:PentPTerSpectra}
\end{figure}

Fig.\ \ref{fig:PentPTerSpectra} presents the results of the
calculations on the three structures of pentacene in
\textit{p}-terphenyl, alongside experimental
data\cite{heinecke_laser_1998,kohler_intersystem_1996}. These are the
absorption spectra that correspond to the data shown in Table 1 in
Section 3.3 of the main text.

\bibliographystyle{unsrt}
\bibliography{EMFTBib}